%% file: main.tex
\documentclass[a4paper,11pt]{article}

\usepackage{jinstpub} 
\usepackage{lineno}
\usepackage{placeins}
\usepackage{soul}
\usepackage{titlesec}
\usepackage{subfigure}
\usepackage[normalem]{ulem}

\title{\boldmath First-Ever Deployment of a SiPM-on-Tile Calorimeter in a Collider: A Parasitic Test with 200 GeV $pp$ Collisions at RHIC. }

\author[a]{Weibin Zhang,}
\author[a]{Sean Preins,}
\author[a]{Jiajun Huang,}
\author[a]{Sebouh J. Paul,}
\author[a]{Ryan Milton,}
\author[a]{Miguel Rodriguez,}
\author[a]{Peter Carney,}
\author[a]{Ryan Tsiao,}
\author[a]{Yousef Abdelkadous,}
\author[a]{Miguel Arratia}

\affiliation[a]{Department of Physics and Astronomy, University of California, Riverside, CA 92521, USA}

\emailAdd{miguel.arratia@ucr.edu}

\abstract{We describe the testing of a prototype SiPM-on-tile iron-scintillator calorimeter at the Relativistic Heavy Ion Collider (RHIC) during its 200~GeV $pp$ run in 2024. The prototype, measuring $20 \times 20 \, \text{cm}^{2}$ and 24 radiation lengths in depth, was positioned in the STAR experimental hall, approximately 8~m from the interaction point and 65~cm from the beam line, covering a pseudorapidity range of about $3.1<\eta<3.4$. By using the dark current of a reference SiPM as a radiation monitor, we estimate that the prototype was exposed to a fluence of about {$10^{10}$ 1-MeV $n_{\mathrm{eq}}$/cm$^2$}. Channel-by-channel calibration was performed in a data-driven way with the signature from minimum-ionizing particles during beam-on conditions. A \textsc{Geant4} detector simulation, with inputs from the \textsc{Pythia8} event generator, describes measurements of energy spectra and hit multiplicities reasonably well. These results mark the first deployment, commissioning, calibration, and long-term operation of a SiPM-on-tile calorimeter in a collider environment. This experimental campaign will guide detector designs and operational strategies for the ePIC detector at the future EIC, as well as other applications.}


\begin{document}

\maketitle
\flushbottom

\input{introduction.tex}
\FloatBarrier
\input{setup.tex}

\FloatBarrier
\newpage 
\input{DAQ}
\input{RadiationMonitoring}
\FloatBarrier
\input{Calibration}
\FloatBarrier
\input{Simulation}

\FloatBarrier
\input{eventselection}
\FloatBarrier
\input{results.tex}

\FloatBarrier
\input{conclusions.tex}

\appendix

\section*{Code and Data Availability}
The \textsc{DD4HEP} simulation model of the prototype can be found in~\cite{zhang_2025_14610981}. The scripts and code used during analysis can be found in~\cite{zhang_2025_14625639}. The data collected by the prototype at RHIC and used in Section~\ref{sec:results} is located at~\cite{zhang_2025_14642181}.

\subsection*{Contributions}
Weibin Zhang was the primary data analyst during data-taking and contributed to the installation and commissioning, and editing of the paper. Sean Preins led the construction of the prototype, performed quality assurance with cosmic rays, and worked on the installation and commissioning. Jiajun Huang led the SiPM-based radiation-monitoring analysis and setup. Sebouh Paul contributed to the detector simulation framework and also CAD-designed the PCB boards and other detector components. Ryan Milton assisted in the development of the detector simulation framework. Miguel Rodriguez contributed to the construction of the prototype and quality assurance. Peter Carney designed and built the hodoscope of the system. Ryan Tsiao, Yousef Abdelkadous, and Peter Carney worked on the installation of the prototype. Miguel Arratia conceived, supervised, and edited this work.

\acknowledgments
Thank you, Bill, we couldn’t have done it without you. Thank you, Thomas, and apologies if we caused you any inconvenience. We thank members of the California EIC consortium, and in particular Oleg Tsai, for valuable feedback related to our design and studies. This work was supported by MRPI program of the University of California Office of the President, award number 00010100. Sean Preins and Ryan Milton were supported by a HEPCAT fellowship from DOE award DE-SC0022313.

\bibliographystyle{utphys} 
\bibliography{biblio.bib}

\end{document}

%% file: introduction.tex
\section{Introduction}
A key advancement in modern calorimetry is the development of SiPM-on-tile technology. This approach integrates SiPMs with plastic scintillator tiles, offering much higher granularity than previous designs. Originally motivated by the concept of particle-flow calorimetry~\cite{Thomson:2009rp} in $e^+e^-$ colliders~\cite{LinearColliderILDConceptGroup-:2010nqx,ILDConceptGroup:2020sfq,Linssen:2012hp,FCC:2018evy,CEPCStudyGroup:2018ghi}, this technology was developed and extensively tested by the CALICE collaboration for more than two decades~\cite{Sefkow:2015hna}. More recently, it has been incorporated into the design of several experiments, including CMS at the High-Luminosity LHC (HL-LHC)~\cite{High-Luminosity:2114693,Contardo:2015bmq,CMS:2017jpq} and ePIC at the future Electron-Ion Collider (EIC)~\cite{Accardi:2012qut}.

While SiPM-on-tile technology was not included in any of the original detector proposals for the EIC~\cite{ATHENA:2022hxb,Adkins:2022jfp,CORE:2022rso}, nor the EIC R\&D programs during the preceding decade~\cite{ULLRICH2022167041,AbdulKhalek:2021gbh}, it has recently been adopted for various subsystems of ePIC. We initially proposed its use in the high granularity insert ($3.0<\eta<4.0$)~\cite{Arratia:2022quz} and it was subsequently adopted for the forward and backward hadronic calorimeters ($1.5<|\eta|<3.0$)~\cite{Bock:2022lwp}, as well as the Zero-Degree Calorimeter ($\eta>6.0$)~\cite{Milton:2024bqv}~\footnote{We also proposed it for the Few-Degree Calorimeter~\cite{Arratia:2023gyx} although it is still not officially included in ePIC.}. These iron-scintillator calorimeters will use nearly one million SiPMs. Their high granularity will enable particle-flow reconstruction and advanced AI/ML techniques, supporting a wide range of physics measurements, such as jets and forward hadrons.

Unlike the latest CALICE~\cite{CALICE:2022uwn} and CMS HGCAL~\cite{CMS:2022jvd} designs, the EIC SiPM-on-tile calorimeters will transfer analog SiPM signals to the rear of the calorimeter via long PCBs, where they will be digitized using ASICs. First proposed in Ref.~\cite{Arratia:2022quz}, this approach minimizes space requirements and prevents ASIC heating within the detector volume, eliminating the need for active cooling.

The EIC SiPM-on-tile calorimeters must withstand a fluence of $10^9$ to $10^{12}$ 1-MeV $n_{\mathrm{eq}}$/cm$^2$ per year, depending on location~\cite{ePICFluence}. Although this fluence creates a ``modest'' damage to SiPMs~\cite{Garutti:2018hfu}, it is exacerbated by room temperature operation. By comparison, the SiPM-on-tile section of the CMS HGCAL will operate at $-30^\circ$C to minimize SiPM dark current and withstand up to $8\times 10^{13}$ 1-MeV $n_{\mathrm{eq}}$/cm$^{2}$ over its lifetime~\cite{CMS:2017jpq}. Considering the impact of temperature on dark current\footnote{The CMS HGCAL technical design report estimates a 30-fold decrease in SiPM dark current~\cite{CMS:2017jpq}.}, the peak fluence of $10^{12}$ 1-MeV $n_{\mathrm{eq}}$/cm$^2$ per year at the EIC is comparable to the lifetime fluence of CMS HGCAL. Both scenarios result in similar SiPM dark currents, and consequently, similar noise levels at their respective operating temperatures. Addressing this challenge requires careful design and operational strategies, such as high-temperature annealing between runs~\cite{Arratia:2022quz}, along with thorough calibration and monitoring.

Testing of SiPM-on-tile technology in the context of EIC detector development remains in its early stages. In 2023, we conducted the first test using a 40-channel, $10 \times 10 \times 40\, \mathrm{cm}^3$ prototype and a 4~GeV positron beam at Jefferson Laboratory~\cite{Arratia:2023xhz}. We also performed benchtop measurements to characterize the timing and light yield of the proposed scintillator tiles and SiPMs~\cite{Arratia:2023rdo}, and studied SiPM radiation damage across the EIC fluence range using 60-MeV protons at the UC Davis Cyclotron~\cite{UCDavisTest}. Other tests carried out by the ePIC collaboration are ongoing or under analysis.

The CALICE collaboration has extensively demonstrated and refined SiPM-on-tile technology through numerous test beams~\cite{Sefkow:2015hna,CALICE:2022uwn,CMS:2022jvd}. However, to our knowledge, no calorimeter prototype has been tested or deployed in a collider environment. Such a test could provide valuable insights into deployment, monitoring, and calibration in harsh radiation environments, informing future detector designs. The Relativistic Heavy Ion Collider (RHIC) at Brookhaven National Laboratory presents an ideal setting for this, as its beam energies and radiation environment are similar to those expected at the EIC, with RHIC's hadron accelerator set to be repurposed for the EIC.

This paper presents results from a test conducted with a prototype deployed in the STAR Hall at RHIC during the 200 GeV proton-proton ($pp$) run from April to October 2024. The goals were to achieve the first deployment of SiPM-on-tile calorimeter technology in a collider, perform in situ calibration of a prototype under realistic conditions, evaluate the impact of radiation damage on detector performance, and benchmark simulations against data.

%% file: setup.tex
\section{Prototype Design and Location in STAR Hall}
\label{sec:setup}
\begin{figure}[h!]
    \centering
        \includegraphics[width=0.99\linewidth]{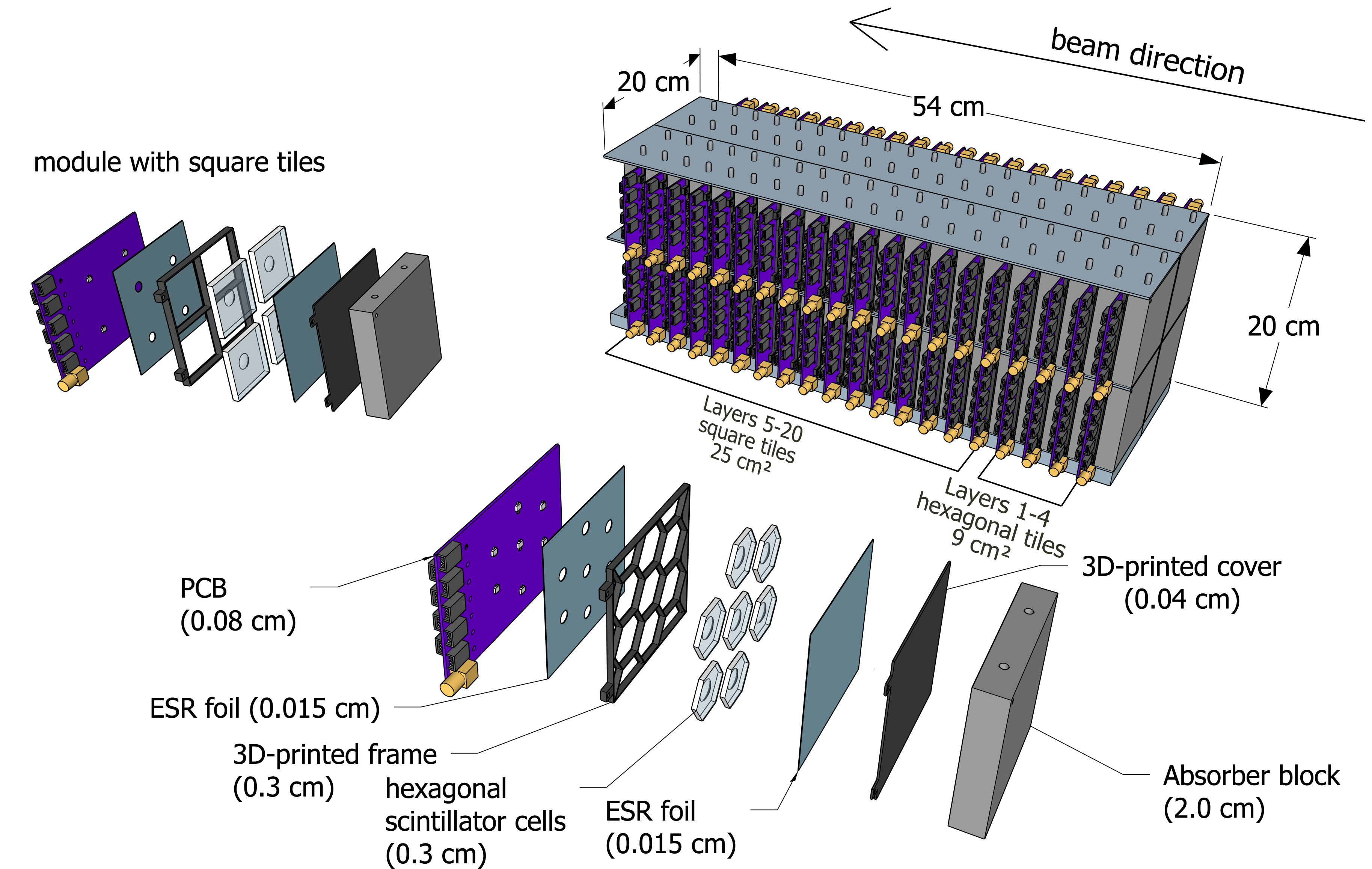}
        \caption{Exploded view of the prototype SiPM-on-tile calorimeter used in this work.}
        \label{fig:prototype}
\end{figure}

Figure~\ref{fig:prototype} shows the design of the SiPM-on-tile calorimeter prototype, which has an area of $19.2\times 19.8$ cm$^2$ and consists of 20 sampling layers. The absorber structure consists of Fe blocks with an area of $9.6\times9.8$ cm$^2$ and a thickness of 2 cm, which corresponds to 1.14 $X_{0}$. The Fe blocks are connected with dowel pins and 2 mm aluminum plates positioned at the middle and top of the prototype, and a $1/2$ in aluminum plate at its base. 

Each active layer contains four quadrants, each defined by the transverse area of an Fe block. The first four and a half layers use 9~cm$^2$ hexagonal scintillating tiles of 3~mm thickness, with seven tiles per quadrant, while the remaining layers up to layer 15 employ 25~cm$^2$ square tiles of 3~mm thickness, with four tiles per quadrant. Layer 15 only has one quadrant, and all layers beyond that are empty. All cells have a dome-shaped cavity (``dimple'') to house the SiPM~\cite{Blazey:2009zz,Simon_2010,Liu:2015cpe}. The SiPMs used in all of layer 3, and one quadrant in layer 4 and 5 are the 1.3~mm Hamamatsu S14160-1315PS. All other layers use 3~mm Hamamatsu S14160-3015PS SiPMs.


The layer-assembly process followed the same procedure as in our earlier prototype tested at Jefferson Lab~\cite{Arratia:2023xhz}. A sheet of ESR foil was first placed on the PCB, which only carries SiPMs and passive components for filtering and a sensor for monitoring temperature~\cite{temperatureSensor}. A 3D-printed frame was then positioned on top of the ESR foil, and the scintillator tiles were fitted into the frame to hold them in place. Another sheet of ESR foil was subsequently added on top, followed by a 3D-printed cover. A subset of the 3D-printed frames was painted with white reflective paint (Saint Gobain BC-621). A pair of bolts were threaded through holes in the SiPM PCB, the 3D-printed frame, and the 3D-printed cover, and secured with nuts to hold all components together. The completed assemblies were then slid into the gaps between the Fe blocks. 

The variety of cell types and SiPM sizes was intentionally diverse for in situ performance testing to further inform our SiPM-on-tile designs~\cite{Arratia:2022quz,Arratia:2023gyx,Milton:2024bqv}. The total number of channels in this prototype was 368, but in the work presented, which corresponds to the 2024 run and the first phase of our experiment, a total of 192 channels were read out, corresponding to the first 9 layers. 

\begin{figure}[h!]
    \centering
    \includegraphics[width=0.75\linewidth]{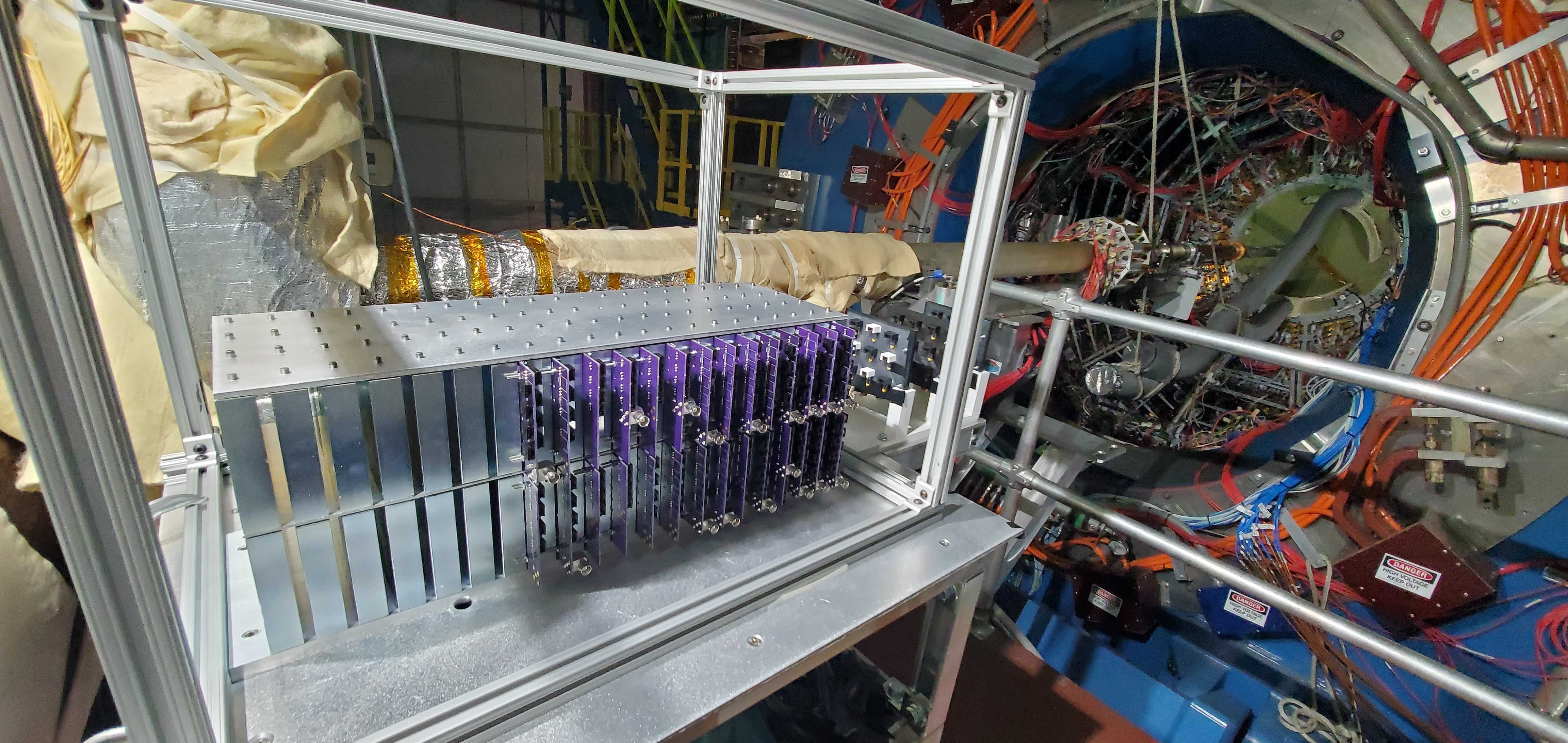}\\
    \includegraphics[width=0.49\linewidth]{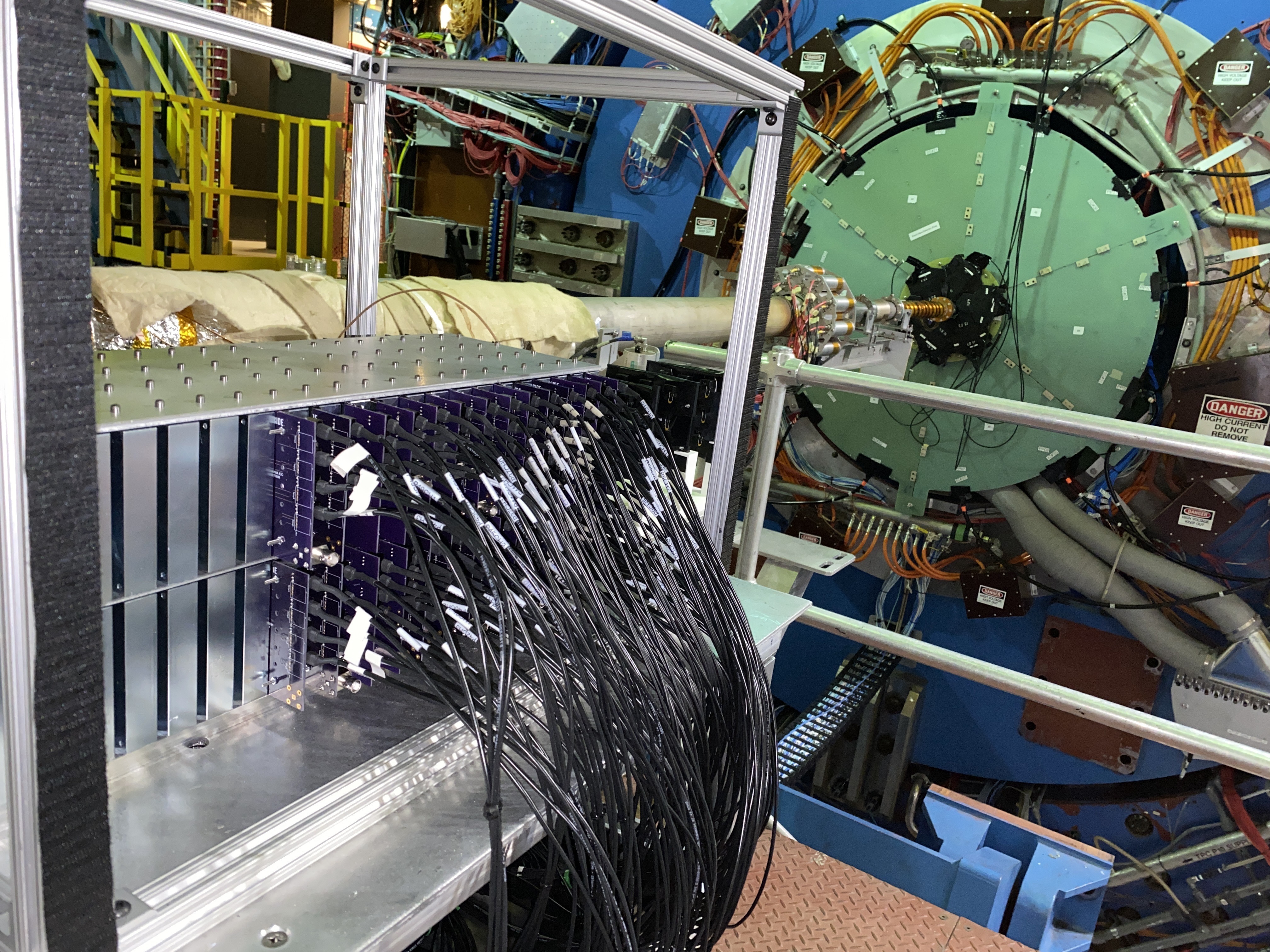}
    \includegraphics[width=0.27\linewidth]{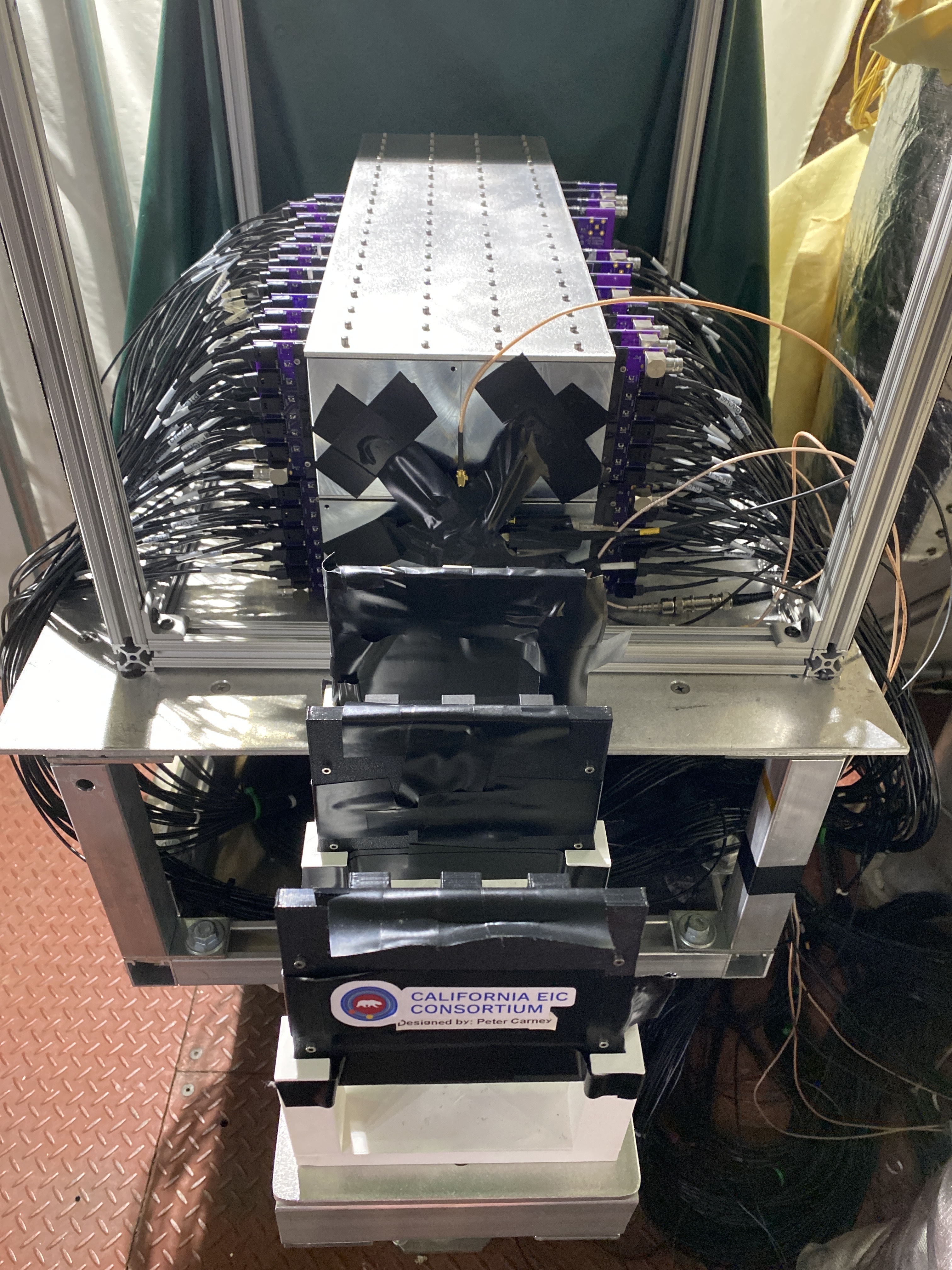}
    \caption{Top: The prototype partially installed on the East Platform of the STAR detector~\cite{STAR:2002eio} (background). The nominal STAR interaction point is visible in this view, with no detector material obstructing the line of sight between the prototype and the interaction point. Bottom left: Rear view of the fully installed prototype. In this view, the STAR detector is in its nominal configuration, with the STAR Event Plane Detector~\cite{Adams:2019fpo} (green disk) being the only material in the line of sight between the prototype and the interaction point.
Bottom right: Front view of the fully installed prototype. The hodoscope detector positioned in front of the prototype was not utilized during the first phase of the experiment presented in this work.
    }
    \label{fig:installation}
\end{figure}

The prototype was positioned on a platform on the east side of the hall that houses the STAR detector~\cite{STAR:2002eio}, as shown in Figure~\ref{fig:installation}. The platform is approximately 8 meters from the interaction point. The prototype's central point was located 65 cm to the left of the beam pipe and aligned vertically with it. The geometrical acceptance of the prototype covers $3.1<\eta<3.4$, overlapping with the nominal acceptance of the high-granularity insert~\cite{Arratia:2022quz} ($3.0<\eta<4.0$). 

The prototype was aligned parallel to the beam pipe, using a non-projective geometry that replicates the intended configuration of the forward and backward endcap detectors for ePIC at the future EIC~\cite{Adkins:2022jfp,ATHENA:2022hxb}.

No detector or support structures obstructed the path between the prototype and the nominal interaction point, except for the STAR Event Plane Detector (EPD)~\cite{Adams:2019fpo}. The STAR EPD consists of a 1.2 cm thick scintillator and a 0.45 cm FR-4 support, which adds up to 0.05 $X_{0}$. The STAR beam pipe is made of a 0.8 mm beryllium layer, which has a negligible impact.

%% file: DAQ.tex
\section{Data Acquisition System and Trigger }
\label{sec:DAQ}
\begin{figure}[h!]
    \centering
    \includegraphics[width=0.7\linewidth]{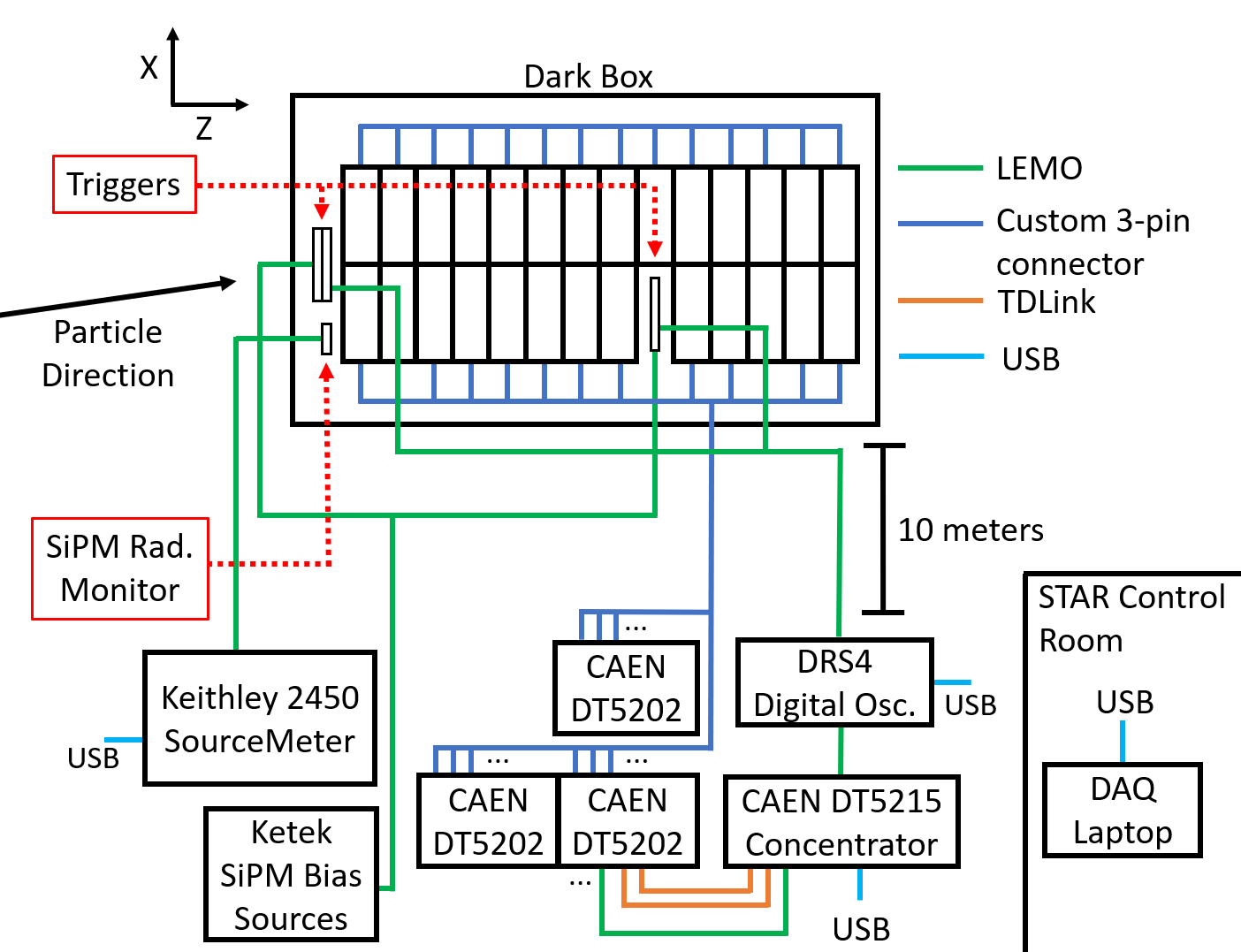}
    \includegraphics[width=0.29\linewidth]{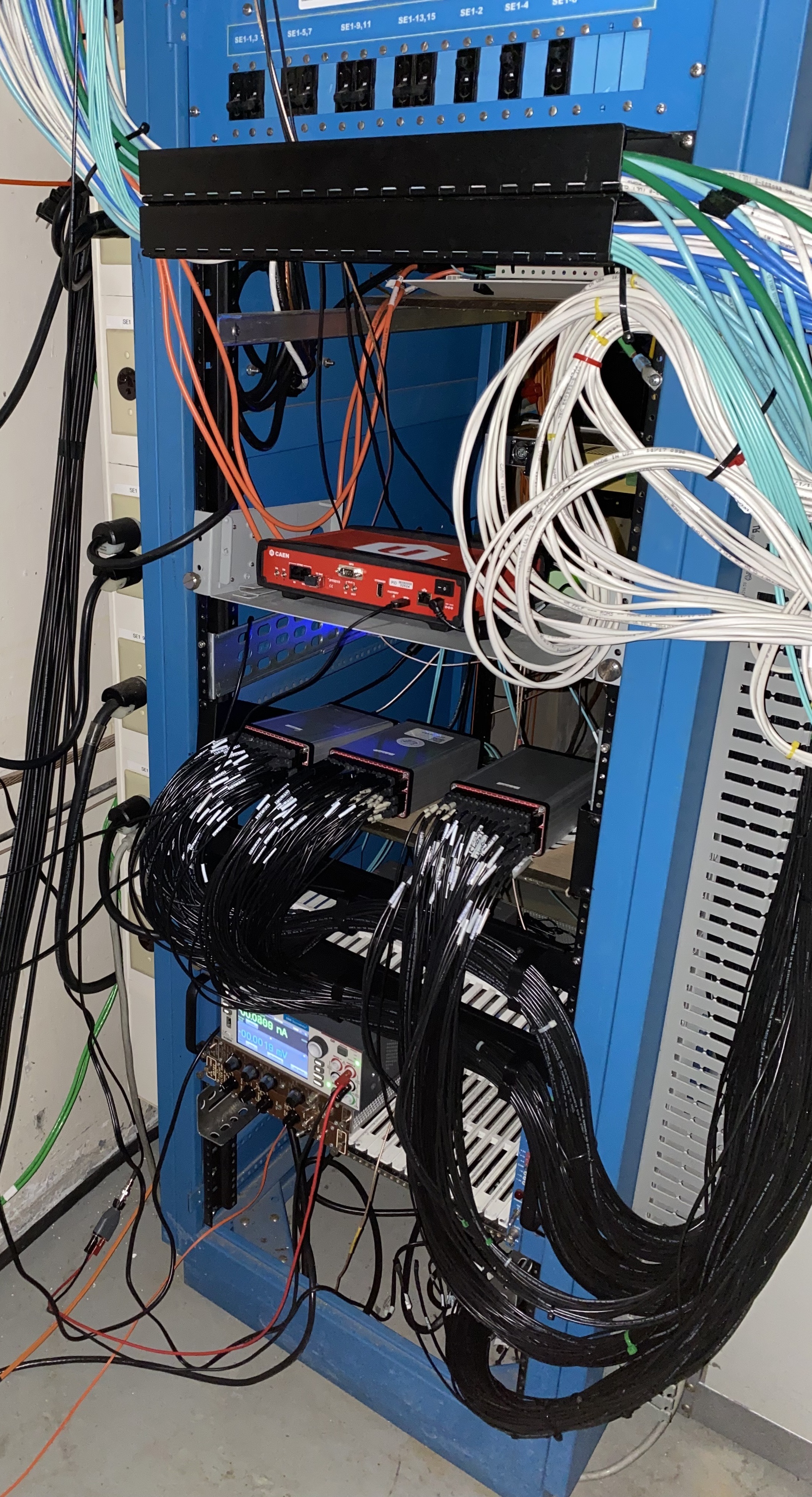}
    \caption{Left: Schematic of the DAQ system and cable connections. Right: Photograph of the rack housing the DAQ system located in the floor of the STAR Hall.} 
    \label{fig:trigger}
\end{figure}

This test was conducted parasitically, without relying on signals from the STAR DAQ system or the RHIC accelerator monitoring system. The standalone trigger, readout, and bias system (referred to as the DAQ system) was composed entirely of off-the-shelf products, described in detail below. The DAQ system was housed in a rack on the STAR Hall floor, approximately 10 meters from the platform that hosts the prototype. Figure~\ref{fig:trigger} shows a diagram and a photograph of the DAQ system.

The DAQ system used signals from three ``trigger tiles'', consisting of plastic scintillators, each measuring $5\times5$ cm$^{2}$ in area and 3 mm in thickness, read out by SENSL SiPMs from the MICRO C series, integrated into SENSL evaluation boards. The scintillators were wrapped in ESR reflective foil, air-coupled to the SENSL boards, and covered with electrical tape. Two trigger tiles, equipped with 6 mm SiPMs, were placed at the front center of the prototype, and one trigger tile with a 3 mm SiPM was positioned in the top right quadrant of the 9th layer. Each SENSL evaluation board was connected via a 10-meter LEMO cable to a dedicated KETEK SiPM Bias Source \cite{KETEK}, which was set to 29 V.

The raw, unamplified analog signals from the three SENSL-SiPM evaluation boards were transmitted via 10-meter LEMO cables to a DRS4 full-waveform digitizer (DRS4 Evaluation Board) \cite{DRS4}. The DRS4 board features four independent input channels that can be sampled simultaneously. The DRS4 sampling frequency was set to 5 GSPS with 1024 sampling points per channel. 

The DRS4 board generated a trigger signal based on configurable logic from the three trigger-tile inputs and variable threshold settings. The maximum readout rate of the DRS4 board is nominally 500 Hz. A waveform generator was used as a random trigger signal for debugging during commissioning. 

The DRS4 connected to a USB hub in the hall, which linked via Ethernet to another USB hub in the STAR main control room, where it connected to a DAQ laptop running Ubuntu 22.04.1. The DRS4 Evaluation Board~\cite{DRS4} was used during beam commissioning and for monitoring under beam-on conditions.

The DRS4 trigger signal served as an external trigger for the CAEN FERS-5200 front-end readout system~\cite{FERS}, which functions as the SiPM readout and bias system for the prototype. This system comprises a CAEN DT5215 FERS concentrator board~\cite{DT5215}, connected in parallel to multiple CAEN DT5202 units~\cite{DT5202} via fiber optics and the CAEN TDLink protocol.

The concentrator board manages multi-board synchronization, slow control, event building, and data readout, and it receives the external trigger signal from the DRS4 board. Each CAEN DT5202 unit provides readout and bias for up to 64 channels through the CITIROC-1A ASIC~\cite{CITIROC}, which features 13-bit analog-to-digital conversion. The CITIROC 1A offers dual-range readout for each channel, enabling adjustable low-gain and high-gain settings. This capability is used in detector calibration, as detailed in Section~\ref{sec:calibration}.

For the data presented in this work, three CAEN DT5202 units were used\footnote{Due to limited availability at the start of the 2024 run, we plan to expand the number of units for the 2025 RHIC heavy-ion run. This expansion will be detailed in a separate publication.}. The CAEN DT5202 units were connected to the SiPM-carrying PCBs via 10-meter CMR-rated single-pair 28 AWG cables with connectors that matched the CAEN A5253 adapter~\cite{A5253}. The concentrator board was connected to the hall USB hub via USB cable.

The CAEN DT5202 units also supplied the bias voltage for the prototype SiPMs. The bias voltage was set to 43~V for all SiPMs, corresponding to an overvoltage of +5~V, assuming a nominal breakdown voltage of 38 V, as specified in their datasheet.

The neutron fluence to which the prototype was exposed was estimated using a single Hamamatsu 14160-3015PS SiPM, positioned in front of the prototype and housed on a custom PCB. The SiPM was connected to a Keithley 2450 SourceMeter~\cite{keithley} via a 10-meter LEMO cable, with the SourceMeter connected to the USB hub in the room. A \textsc{LabVIEW} script periodically performed IV measurements of the SiPM.

As detailed in Section~\ref{sec:IV}, the dark current from this SiPM was compared to reference data from a benchmark dataset~\cite{UCDavisTest} to estimate the integrated fluence received by the reference SiPM, and by extension, the prototype. 

The dark current of the other SiPMs was monitored at a lower resolution using the CAEN DT5202 units. However, these measurements lacked the precision of the Keithley 2450, as the DT5202 integrates the total current from all SiPMs in the prototype. 

The prototype was operated at room temperature (approximately 24$^\circ$C) in the STAR Hall. No adjustments were made to the SiPM bias voltage to account for temperature fluctuations, which were minimal due to the temperature-controlled environment in the STAR Hall, as confirmed by our temperature sensor.

Data collection was managed using CAEN Janus DAQ software (version 3.6.0), enabling automated operation~\footnote{The Janus software configuration was structured as follows: Runs were initiated asynchronously (StartRunMode: ASYNC) and stopped manually (StopRunMode: MANUAL). The run duration was set to 90 minutes (PresetTime: 90 m). The system operated in spectroscopy mode (AcquisitionMode: SPECTROSCOPY) with ToT enabled (EnableToT: 1) and used T0-IN as the bunch trigger source (BunchTrgSource: T0-IN). The fast shaper input was set to high-gain preamplifiers (FastShaperInput: HG-PA), with coarse thresholds of 260 for timing discriminators (TD\_CoarseThreshold: 260) and 250 for charge discriminators (QD\_CoarseThreshold: 250). In spectroscopy mode, both high-gain and low-gain outputs were recorded (GainSelect: BOTH), with gains of 55 and 27, respectively (HG\_Gain: 55, LG\_Gain: 27). The pedestal was set at 160 ADC (Pedestal: 160), with shaping times of 25 ns for both outputs (HG\_ShapingTime: 25 ns, LG\_ShapingTime: 25 ns) and a hold delay of 100 ns (HoldDelay: 100 ns).}. Each run lasted 90 minutes, and upon completion, the data was automatically added to a custom SQLite database, analyzed using the \textsc{ROOT} framework~\cite{Brun:1997pa}, and updated in real time on our dedicated website~\cite{webcali}. Additionally, pedestal and MIP runs were performed manually\footnote{Within Janus, pedestal and calibration runs were executed by setting the CAEN Janus parameter \textsc{TrefSource} to \textsc{PTRG} and \textsc{TLOGIC}, respectively}. Pedestal runs were collected daily, while MIP runs were conducted periodically.

The trigger thresholds for various trigger logic configurations (AND/OR) were set between 10 and 50 mV, resulting in rates from a few Hz to several tens of Hz under typical conditions.

The official RHIC 2024 run of 200 GeV $pp$ collisions took place from April through the end of September, followed by a two-week period of heavy-ion ($AuAu$) collisions in October. The commissioning period for our test, primarily focused on testing and debugging various aspects of the DAQ software, lasted until June 4th.   

The events collected by this DAQ system include real $pp$ collisions, along with particles from other sources such as beam-gas and beam-halo interactions. In the absence of reference signals from other detectors or a reference clock from RHIC or STAR, the setup configuration and trigger strategy were unable to reject these background events.

%% file: RadiationMonitoring.tex
\section{Radiation Fluence Monitoring with Reference SiPM}
\label{sec:IV}
One of the key parameters monitored throughout the run was the radiation fluence. This was measured using the reference SiPM positioned in front of the prototype, as described in Section~\ref{sec:DAQ}. The SiPM's dark current, which increases linearly with fluence in the relevant range~\cite{UCDavisTest}, served as an indicator of the radiation fluence to which the prototype was exposed. 

\begin{figure}[h!]
    \centering
    \includegraphics[width=0.49\linewidth]{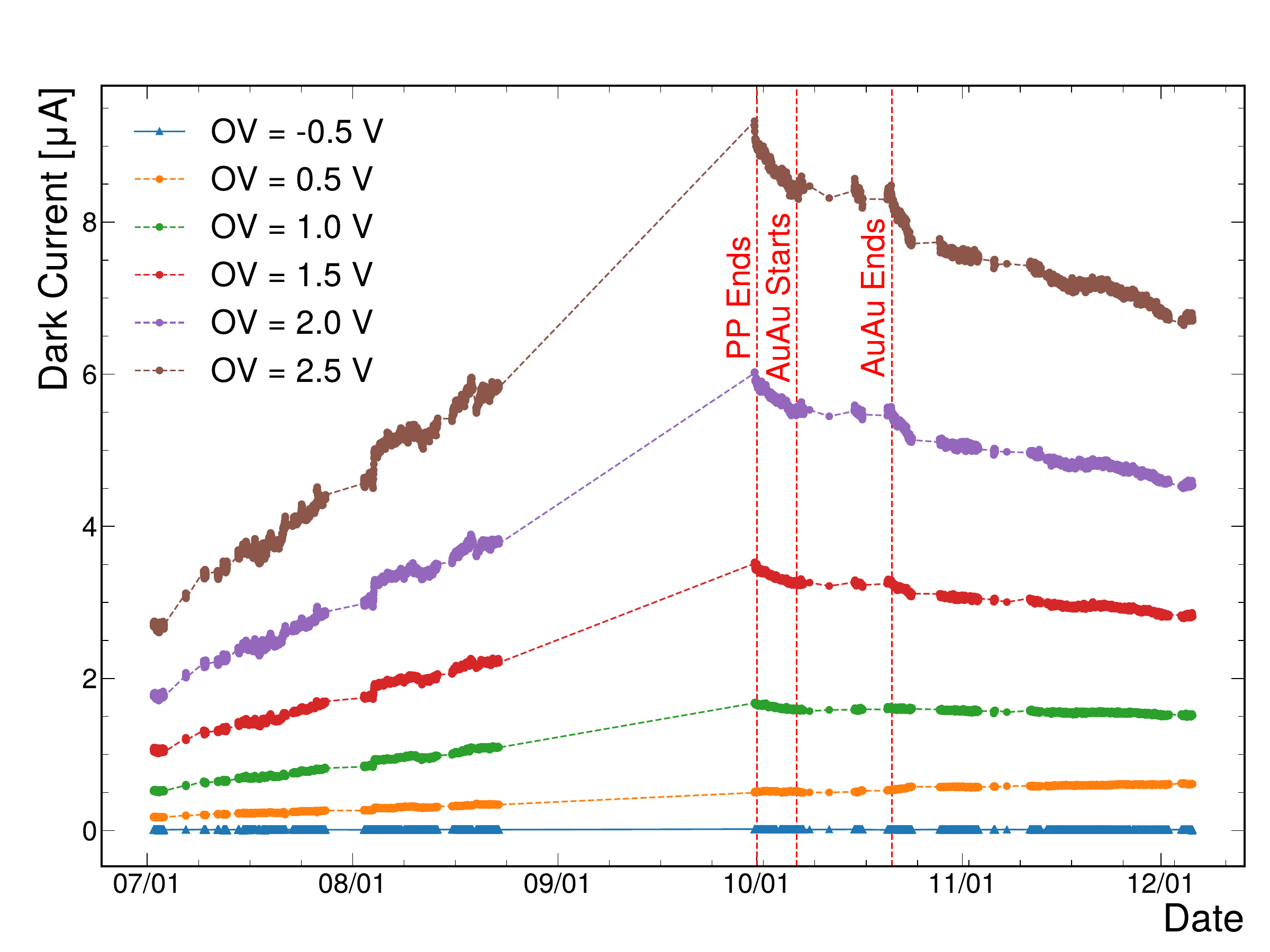}
    \includegraphics[width=0.49\linewidth]{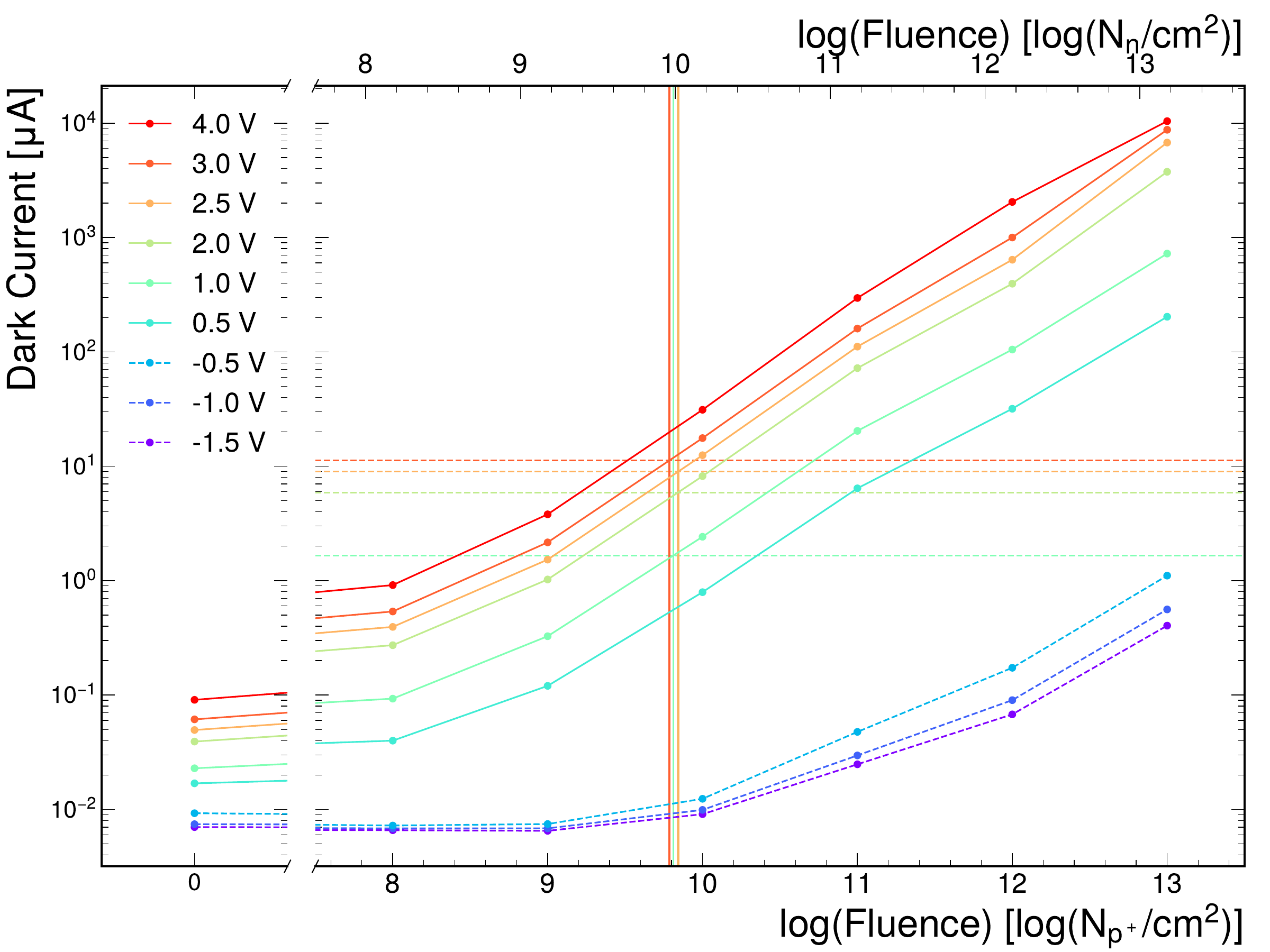}
    \caption{Left: Dark current of the reference SiPM (model Hamamatsu S14160-3015PS) as a function of time. A loss of connection with the Keithley 2450 unit occurred for approximately 40 days, from mid-August to late September, leading to the large gap in the plot. Right: Results of the benchmark data~\cite{UCDavisTest}. The horizontal lines represent the dark current measured on September 30th, with the cross points indicating the inferred peak radiation level on that date. Consistent results were observed across different breakdown voltages.}
    \label{fig:radiation}
\end{figure}

Figure~\ref{fig:radiation} shows the measured dark current at different bias voltages over time. The first measurements were recorded on July 1st, following a longer commissioning period required for this monitoring system. When operated at $+2$ V overvoltage, the reference SiPM drew 2 $\mu$A during the initial measurement, increasing to 4 $\mu$A by August 14th. Data taking was then interrupted due to a software issue and resumed shortly after the conclusion of the RHIC $pp$ run on September 30th, at which point the SiPM drew approximately 6 $\mu$A. For reference, measurements of an unirradiated SiPM of the same model yielded 0.04 $\mu$A when biased at $+$2 V overvoltage.

Room-temperature annealing effects were observed during the beam-off week from October 1st to 7th and immediately after the conclusion of Run 2024, as evidenced by the drop in dark current during these periods. 

The integrated radiation fluence for the reference SiPM was quantified using a benchmark calibration dataset, where the response of SiPMs of the same model was measured at various radiation fluences. This calibration was conducted by irradiating SiPMs with 60 MeV proton beams at the UC Davis McClellan Nuclear Research Center’s Cyclotron facility. In particular, the dark current measured at the end of the $pp$ run, compared to the reference dataset of 6 $\mu$A at $+$2V, yields an estimated fluence of approximately $10^{9.8}$ p$^+$/cm$^2$. The dark current measured at various overvoltages was also used to infer the fluence, yielding consistent results.

This is illustrated in the right panel of Figure~\ref{fig:radiation}, where the measured current at various bias voltages is shown as horizontal lines, and the reference dataset~\cite{UCDavisTest} is represented by markers with interpolating lines. The horizontal lines and the corresponding reference curves intersect at around $10^{9.8}$ p$^+$/cm$^2$. Applying a factor of 1.5 to account for the difference in damage factors between 60 MeV protons and 1 MeV neutrons on silicon, this fluence is equivalent to $10^{10}$ 1 MeV $n_{eq}$/cm$^2$. A systematic uncertainty of $\pm$20$\%$ is assigned on this estimated fluence, based on uncertainties in the reference dataset and the effects of room-temperature annealing that are not accounted for.

While the total radiation dose that the prototype received is not much less than the peak fluence expected at the EIC when running at peak luminosity, a large portion of the forward hadronic calorimeter will receive a total radiation dose that is smaller than or similar to what this prototype received~\cite{ePICFluence}. 

%% file: Calibration.tex
\section{Calibration}
\label{sec:calibration}
The detector calibration was performed channel-by-channel multiple times throughout the $pp$ run, using the signature of minimum-ionizing particles. The high-gain setting of the CAEN DT5202 unit was selected to ensure clear separation between the pedestal and the MIP peaks.

Unlike previous beam tests that used dedicated cosmic-ray runs~\cite{Arratia:2023xhz} or muon beams~\cite{CALICE:2022uwn}, the MIP calibration in this work was performed using beam-on data. This approach enabled much faster collection of calibration data compared to dedicated cosmic-ray runs, which could not be reliably conducted due to the absence of scheduled stops during the RHIC 2024 run.

\begin{figure}[h!]
    \centering
    \includegraphics[width=0.49\linewidth]{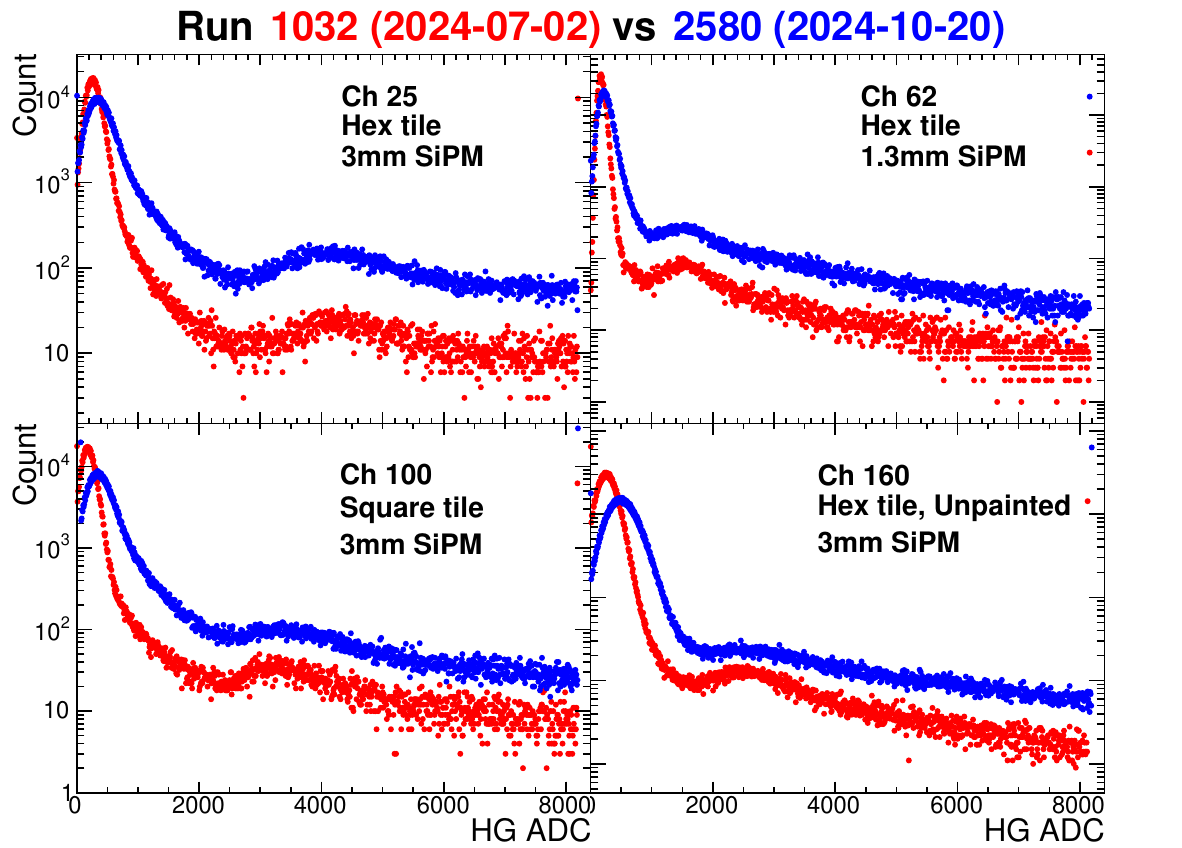}
    \includegraphics[width=0.49\linewidth]{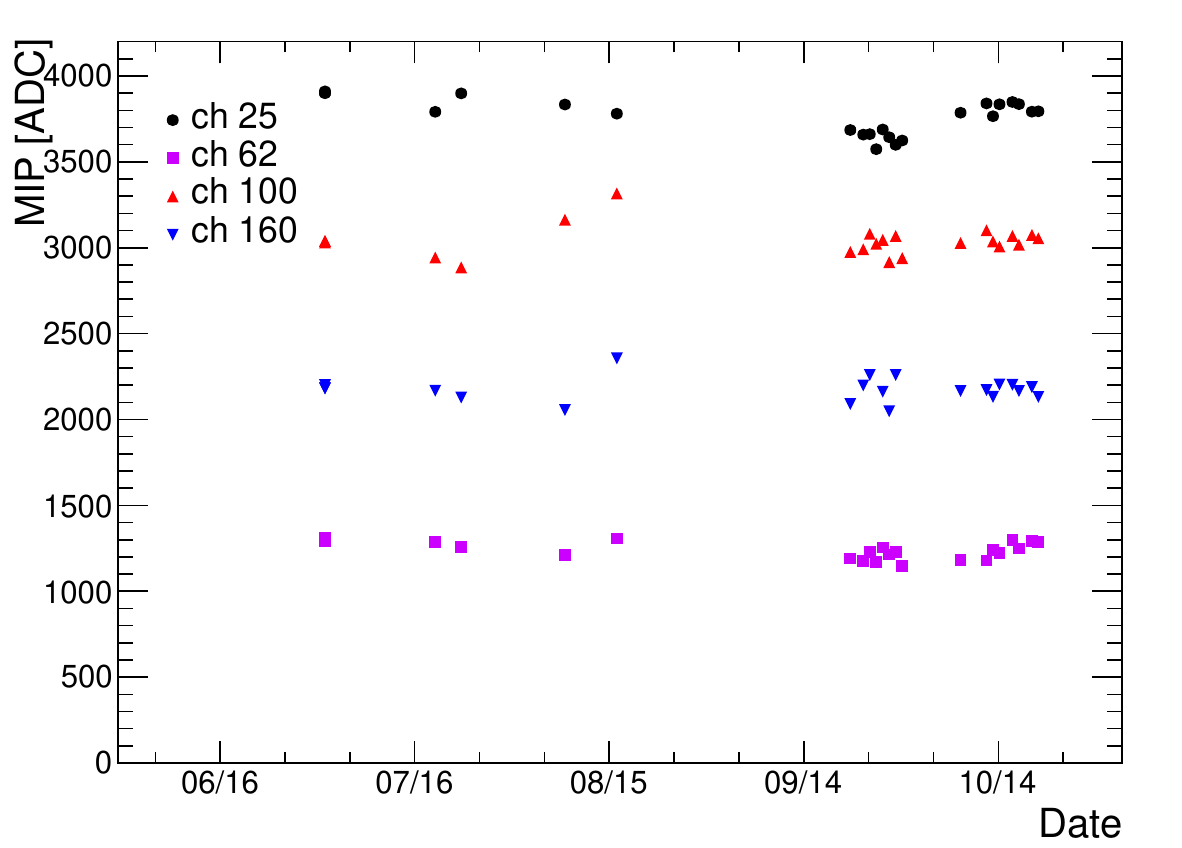}
  \caption{Left: Comparison of high-gain ADC spectra for representative channels of different types between early and late MIP calibration runs. All distributions are normalized to unity. The first peak in each spectrum represents the pedestal, while the second peak corresponds to the MIP. The mean value and width of the pedestal peak increase over time due to higher SiPM dark current resulting from radiation damage.
Right: MIP calibration values for the representative channels over time.}
    \label{fig:channel-MIP-calibration}
\end{figure}

During beam-on data collection, the MIP signature appeared in the high-gain readout as a Landau peak superimposed on a monotonically decreasing hit-energy spectrum (in ADCs). This is illustrated in the left panel of Figure~\ref{fig:channel-MIP-calibration}, which shows four different types of channels from two representative runs. As expected, the pedestal was noticeably wider in the later runs due to increased SiPM dark current. However, the MIP signal remained clearly separated from the pedestal peak in all cases, as further quantified later in this section.

The high-gain MIP calibration values were calculated as the difference between the MIP peak, determined using a peak-finding method with smoothing, and the pedestal peak, obtained through a Gaussian fit. This fit was performed on a run-by-run basis to account for pedestal broadening and mean shifts caused by increased SiPM dark current over time.

The right panel of Figure~\ref{fig:channel-MIP-calibration} shows the MIP calibration value for four representative channels as a function of time. The MIP calibration remained stable within $\pm10\%$ from the average for all cases. Sources of variation in the calibration values for a given channel may include instabilities in the readout electronics, luminosity-dependent backgrounds, or biases in the peak extraction method. 

\begin{figure}[h!]
    \centering
    \includegraphics[width=0.98\linewidth]{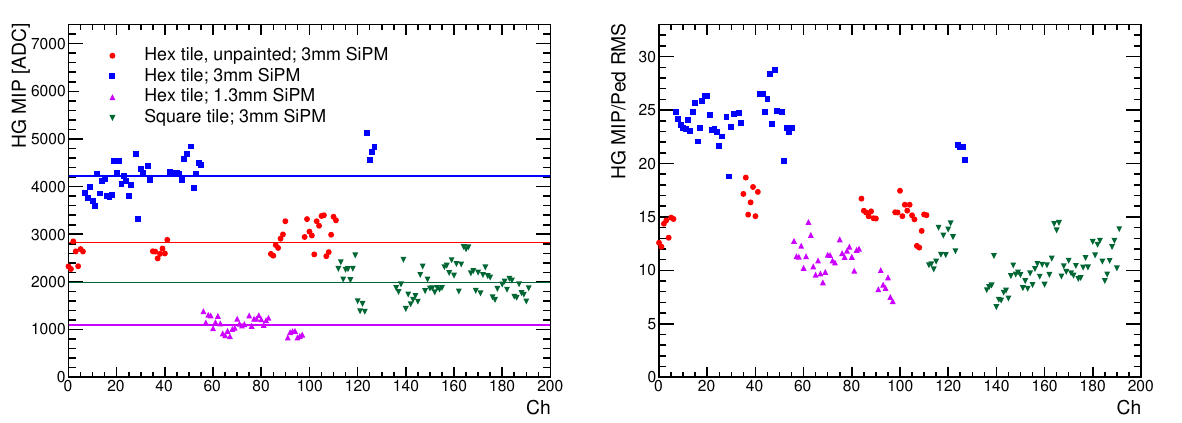}
    \caption{
    Left: Channel-by-channel MIP calibration in run 2580 (10/20/2024). Four sets of MIP values are observed, corresponding to four different combinations of scintillating tiles and SiPM sizes. Horizontal lines represent the average MIP values for each set.
    Right: Ratio of MIP values to the pedestal width per channel.
    }
    \label{fig:HG-MIP}
\end{figure}

Due to variations in scintillating tile shapes, SiPM sizes, and reflective paint treatment, four distinct sets of MIP values are expected. Figure~\ref{fig:HG-MIP} shows four distinct groupings corresponding to these different cases. The channel-by-channel variation in MIP calibration for each category, which ranges approximately $\pm$25\% from the average, can be attributed to misalignment between the tile and SiPM~\cite{deSilva:2020mak}, variations in SiPM breakdown voltage, or other manufacturing factors. 

The right panel of Figure~\ref{fig:HG-MIP} shows the ratio of MIP values to their corresponding pedestal widths during a run near the end of the 2024 RHIC $pp$ period. In all cases, the ratio exceeds 6, indicating a clear separation between pedestal and MIP peaks even after the prototype had been exposed to approximately $10^{10}$ 1 MeV $n_{eq}$/cm$^2$, as discussed in Section~\ref{sec:IV}. 

\begin{figure}
    \centering
    \includegraphics[width=0.5\linewidth]{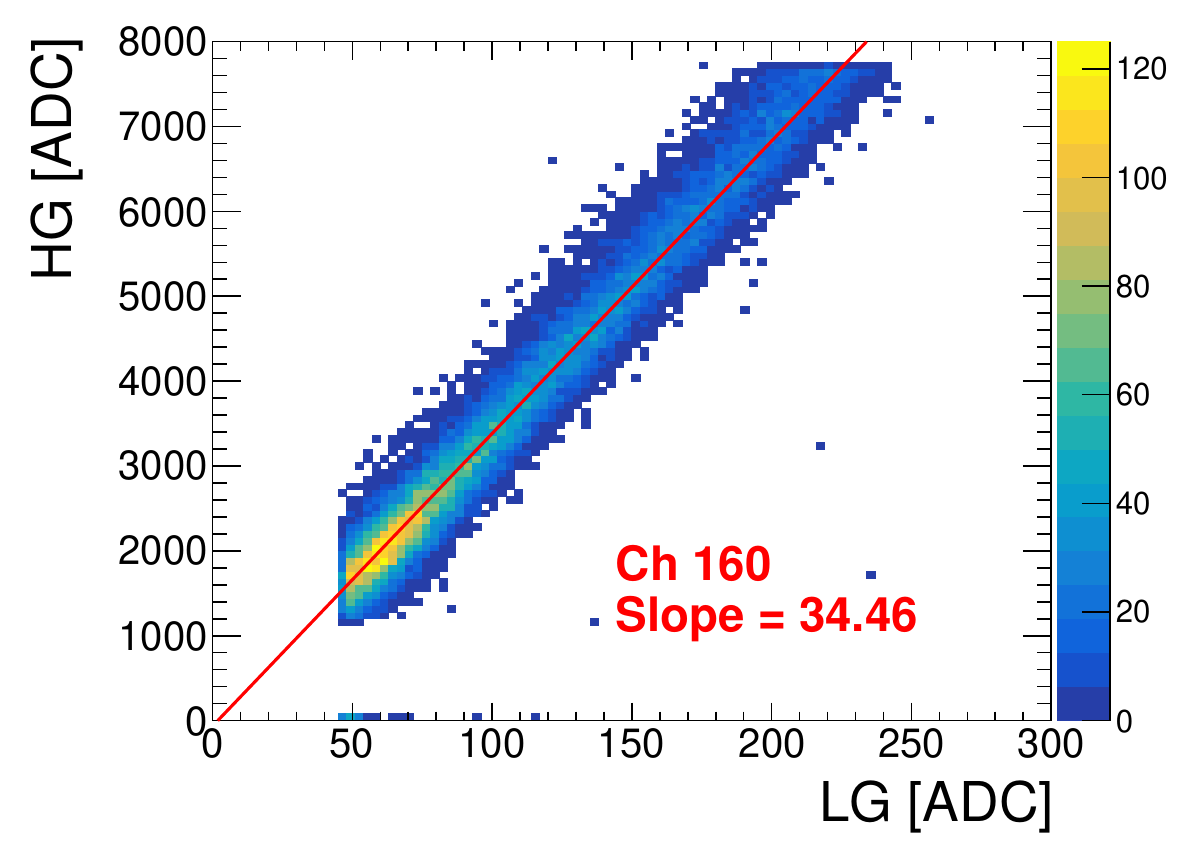}
    \caption{Correlation between high-gain and low-gain reading for a representative channel during run 2019 (09/16/2024). The red line represents a linear fit.}
    \label{fig:HG2LG}
\end{figure}

The MIP peak is not visible in the low-gain readout. Low-gain calibration values were inferred from the high-gain MIP value and a proportionality constant obtained from dual-range readings of beam-on data in the overlapping region, where the high-gain readout was not yet saturated. Figure~\ref{fig:HG2LG} shows an example of the correlation between low-gain and high-gain settings for a representative run and channel. A linear fit was performed to obtain the calibration factor.  

The low-gain MIP values were then used to calibrate the low-gain readings for each channel as follows. Channel-by-channel raw energy measurements, in ADC units, were pedestal-subtracted and calibrated to the MIP energy scale. The pedestal mean and width for each channel were determined by performing a Gaussian fit for each run, using dedicated random-trigger pedestal runs\footnote{The pedestal widths were observed to increase in a manner consistent with being proportional to the square root of the fluence (dark current of the monitor SiPM), as expected.}. The pedestal-subtracted and calibrated hit energy for the $i$-th channel was then:
\begin{equation}                                                                
    E_{i} [\mathrm{MIP}]  =  \frac{\left ( E_{i} [\mathrm{ADC}] - \langle\mathrm{P}\rangle_{i} [\mathrm{ADC}] \right )}{ \mathrm{C}_{i} [\mathrm{ADC/MIP}]}
    \label{eq:calibration}
\end{equation}
Here, the mean pedestal $\langle\mathrm{P}\rangle_{i}$ and calibration values $C_{i}$ were determined for each run from the nearest pedestal and MIP calibration runs. 

A total of 180 channels were successfully calibrated throughout the entire 2024 RHIC run, representing 94\% of all readout channels. Channels that were not successfully calibrated were either dead, likely due to connection failures, or exhibited anomalously high rates and were masked.

A systematic uncertainty of $\pm 10\%$ is assigned to the MIP energy scale, which is assumed to be fully correlated across all channels. This uncertainty is estimated by varying the peak extraction method and accounting for possible variations in beam conditions used for the MIP calibration.

%% file: Simulation.tex
\section{Simulation}
\label{sec:simulation}
We simulated the prototype using \textsc{Geant4}~\cite{GEANT4:2002zbu} (v11.02.p2) with the FTFP\_BERT physics list. The simulation was carried out using the DD4HEP framework~\cite{Frank:2014zya}\footnote{We used this framework in prior studies~\cite{Milton:2024bqv,Arratia:2022quz,Paul:2024fww,Acosta:2023nuw,Acosta:2023zik} and validated it by benchmarking against CALICE test-beam data~\cite{CALICE:2012eac,CALICE:2022uwn} and previous test beams at Jefferson Lab~\cite{Arratia:2023xhz}.}. The prototype geometry and materials were modeled according to the descriptions provided in Section~\ref{sec:setup}. The material from the STAR Event Plane Detector was incorporated into the simulation. No magnetic field was included in the simulations because the 0.5 T solenoid field from STAR is expected to have a negligible effect in the pseudorapidity range of interest. The trigger tiles described in Section~\ref{sec:setup} were also included.

We digitized the hit energies with a dynamic range of 200 MeV and a 13-bit ADC. We set the pedestal values in the simulation to 150 ADC with a Gaussian width of 7.5 ADCs, reflecting values representative of those observed in data under low-gain settings.
Dead channels were also accounted for in the simulation chain. A high-energy muon simulation was used to calibrate the reconstructed hit energy to the MIP scale, following Equation~\ref{eq:calibration}.

\begin{figure}[h!]
    \centering
    \includegraphics[width=0.49\linewidth]{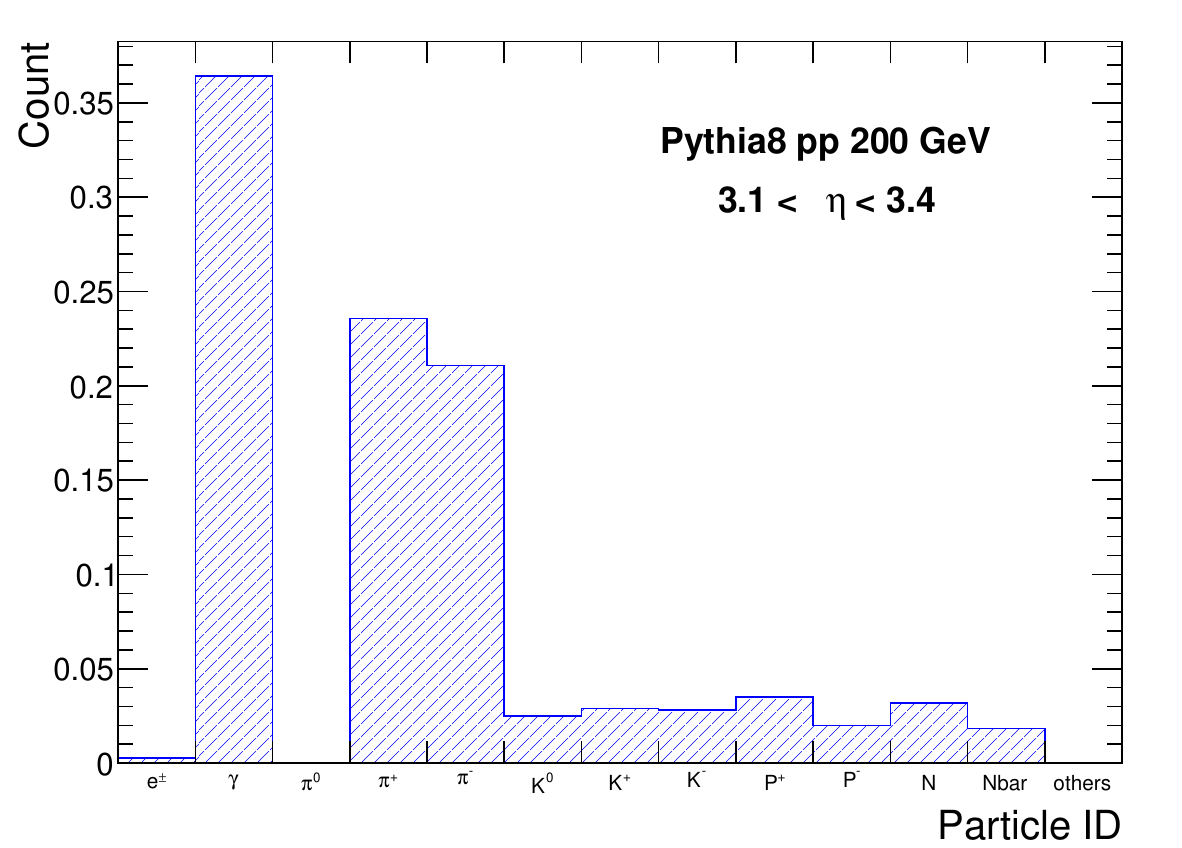}
    \includegraphics[width=0.49\linewidth]{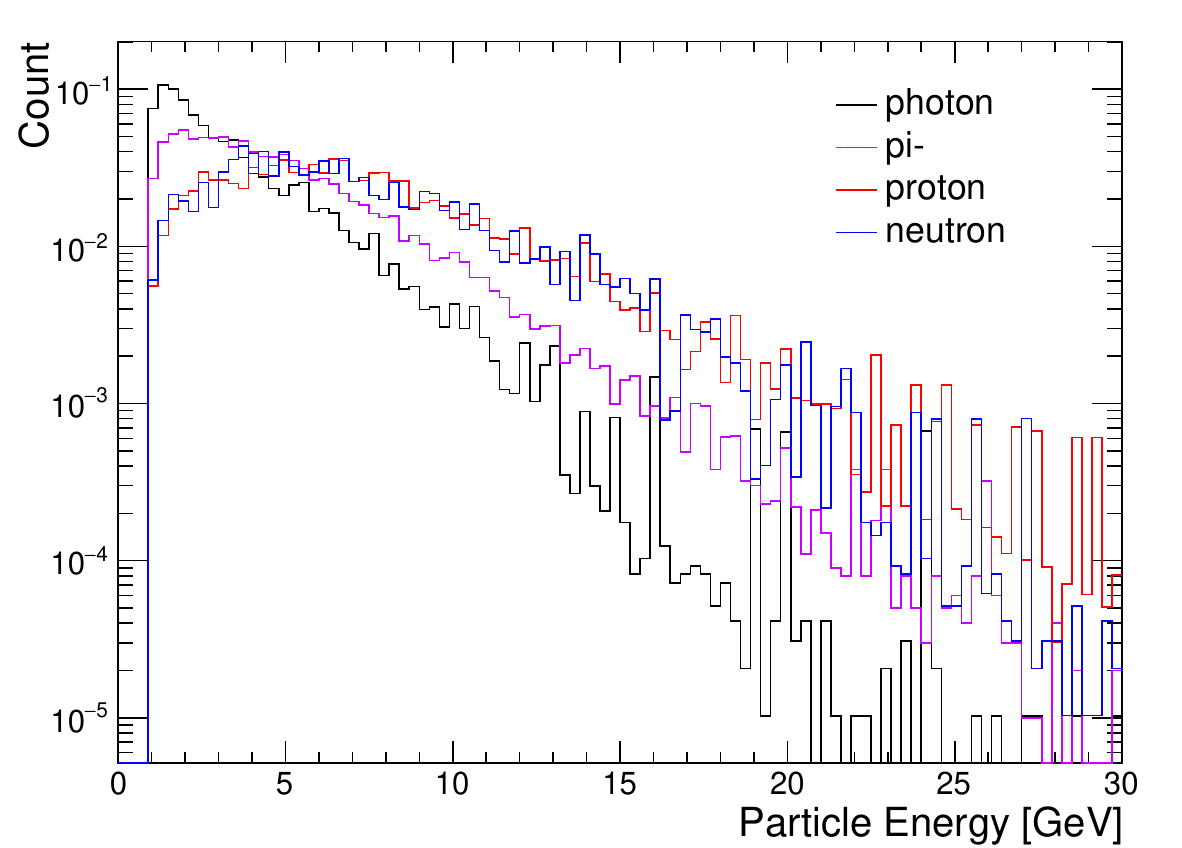}
  \caption{Particle composition (left) and energy spectra (right) expected at the prototype location from truth-level \textsc{Pythia8} events passing the geometrical filter. Both plots are normalized to unity.}        
    \label{fig:pythia8}
\end{figure}

We used the Monte Carlo event generator \textsc{Pythia8} (version 8.307)~\cite{Sjostrand:2014zea} with the default Monash 2013 tune to simulate $pp$ collisions at 200 GeV as input to the \textsc{Geant4} simulation chain. All hard QCD processes were enabled, and no minimum $p_T$ requirement was applied to generate minimum-bias $pp$ events. A geometric event filter was applied, requiring at least one final-state particle to reach the prototype by projecting their trajectories assuming no magnetic field. The resulting particle composition is shown in the left panel of Figure~\ref{fig:pythia8}. It is predominantly composed of charged pions and photons from neutral pion decay, along with some hadrons such as protons and neutrons. The corresponding energy spectra are shown in the right panel of Figure~\ref{fig:pythia8}.

Studies with single-photon simulations indicate a sampling fraction of about 1$\%$, with 1 GeV corresponding to approximately 30 MIP. This MIP-to-GeV factor is not used in the analysis. Instead, results are presented on the MIP scale for both data and simulation.



%% file: eventselection.tex
\section{Trigger, Hit, and Event selection }
\label{sec:selection}
Events were selected using a trigger based on either a logical OR using the front trigger tiles T1 or T2 (referred to as T1$\vee$T2) or a signal from the middle trigger tile T3 (referred to as T3). In both cases, the threshold corresponds to 3 MIP.

Hits with energy in the range $0.5 < E < 140$ MIP were retained for further analysis. The lower threshold of 0.5 MIP minimizes the contribution of fake hits from pedestal fluctuations, as it corresponds to at least three times the pedestal width, as shown in Figure~\ref{fig:HG-MIP}, and is consistent with previous CALICE studies~\cite{CALICE:2022uwn}. The upper limit of 140 MIP was chosen to mitigate saturation effects due to the 13-bit ADC capacity of the readout used. It also renders negligible the non-linearity of the SiPM response caused by high pixel occupancy.

The total energy of all hits passing the energy threshold defines the event energy. Events with energies exceeding 150 MIP (5 GeV) were selected for further analysis. A total of 40,079 T1$\vee$T2 events and 93,970 T3 events passed the selection.

The same hit, event, and trigger selection criteria applied to the data were also applied to the simulation. A total of 55,609 T1$\vee$T2 and 41,139 T3 simulated events passed the selection.

%% file: results.tex
\section{Results}
\label{sec:results}
\begin{figure}[h!]
    \centering
    \includegraphics[width=0.49\linewidth]{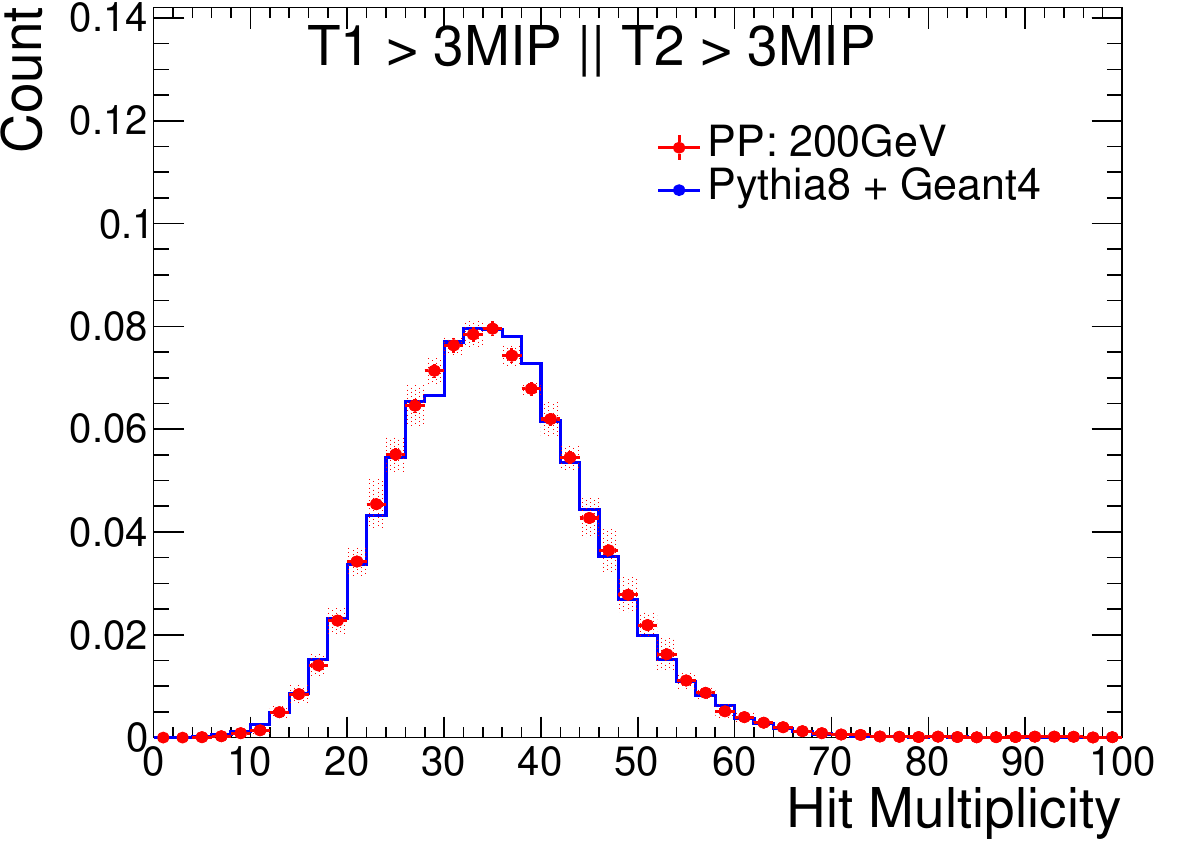}
    \includegraphics[width=0.49\linewidth]{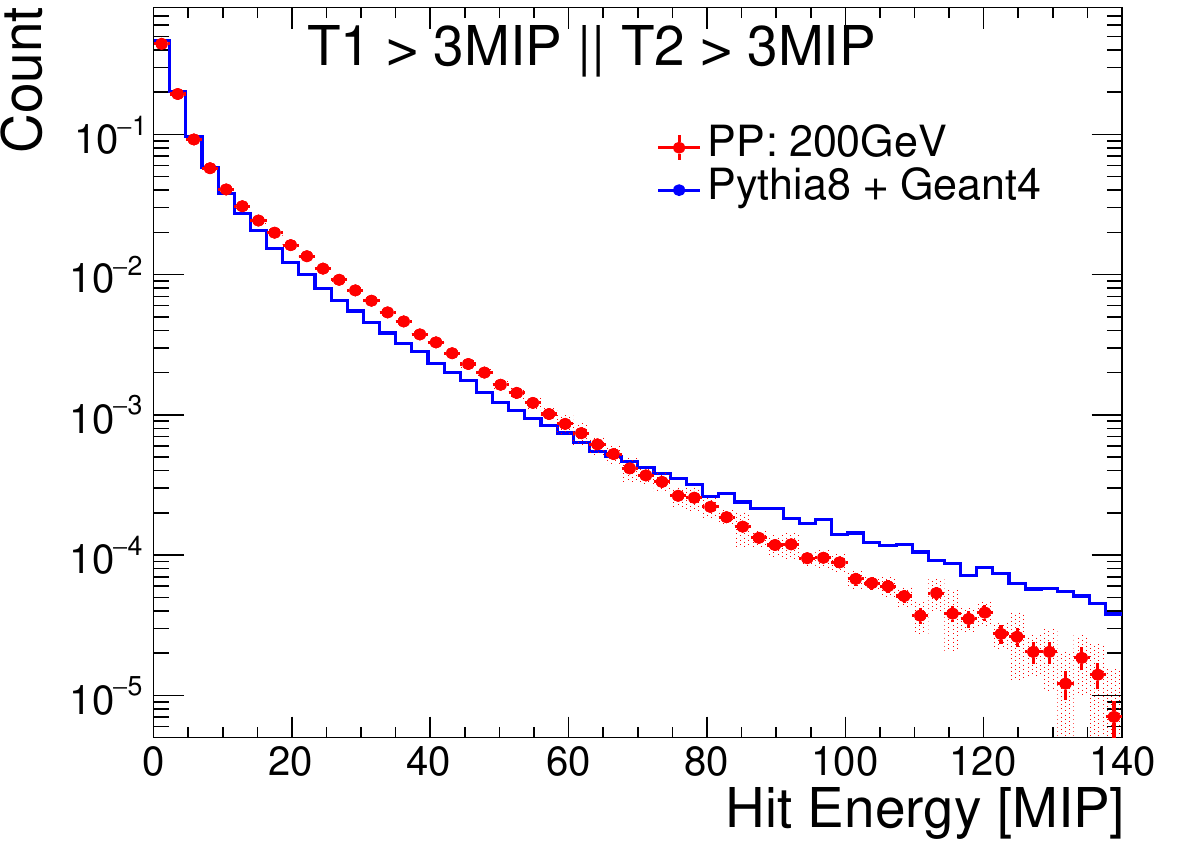}\\
    \includegraphics[width=0.49\linewidth]{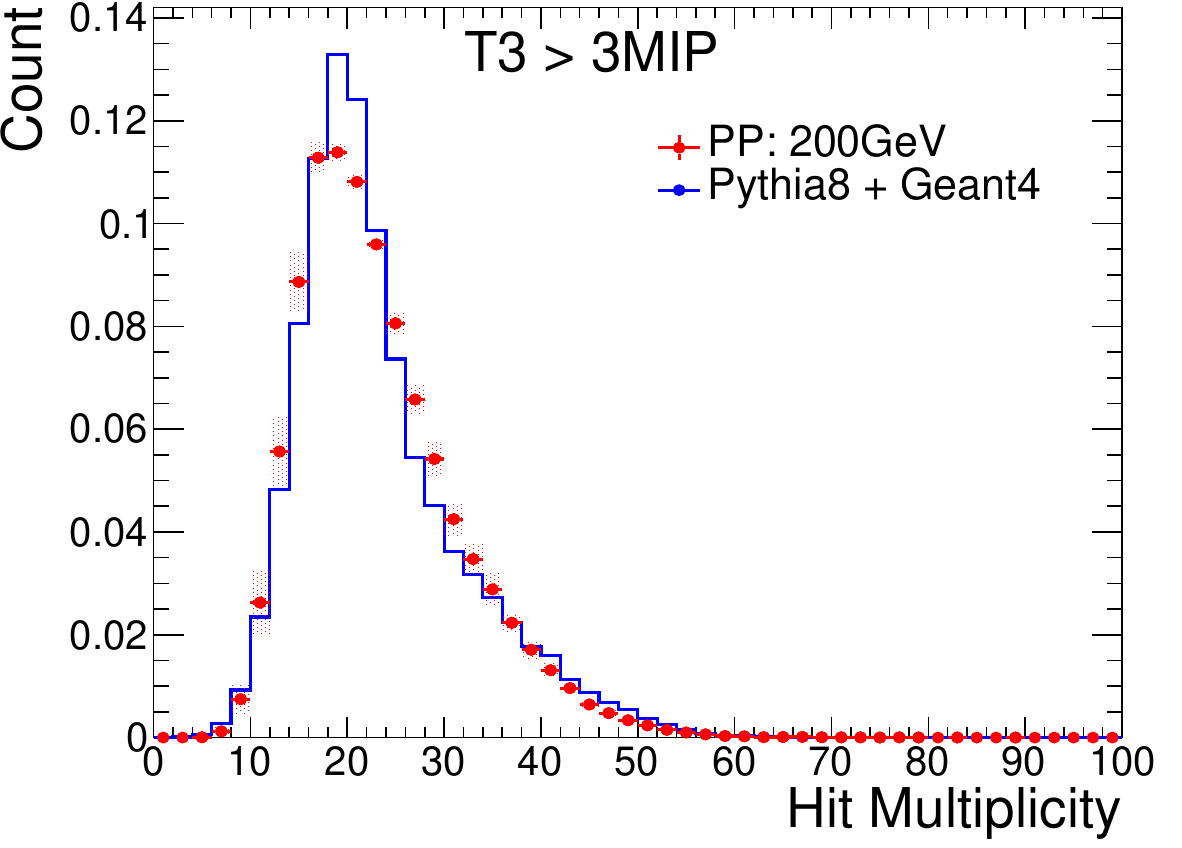}
    \includegraphics[width=0.49\linewidth]{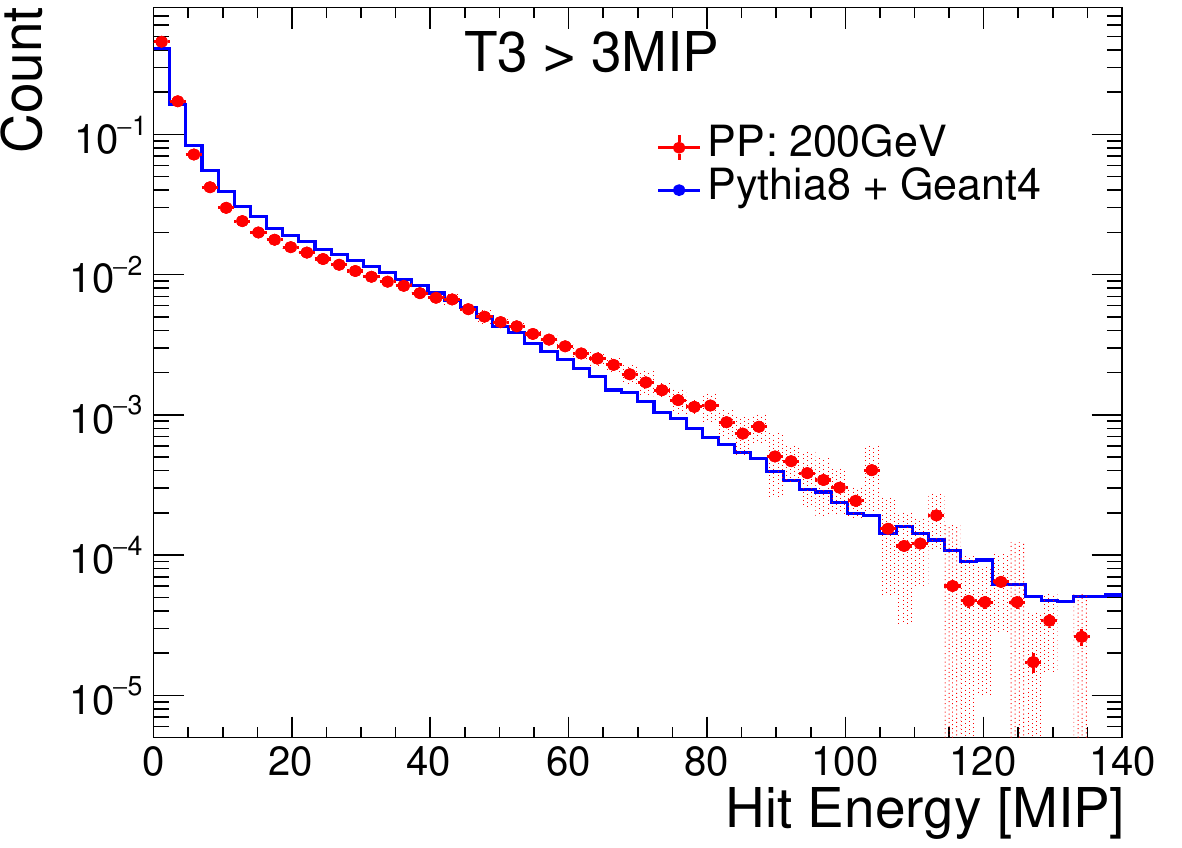}
    \caption{Comparison of data and simulation for hit multiplicity and hit energy spectra for the T1$\vee$T2 trigger (top row) and the T3 trigger (bottom row). All distributions are normalized to unity. Error bars and bands represent statistical and systematic uncertainties, respectively.}
    \label{fig:hit-comparison}
\end{figure}

Figure~\ref{fig:hit-comparison} compares data and simulation for hit multiplicity and hit energy spectra for both T1$\vee$T2 and T3 triggers. All distributions are normalized to unity. T1$\vee$T2 events exhibit a Gaussian-like distribution with an average of approximately 30 hits, while T3 events peak at around 20 hits and display a more pronounced right tail.

For hit energy, low-energy hits ($<$10 MIP) dominate in both cases. However, in the higher energy region ($>$10 MIP), T1$\vee$T2 events show a steeper power-law decline compared to T3 events. The difference between T1$\vee$T2 and T3 events arises from the different trigger locations and energy sampling of the events, which is reasonably well described by the simulation.
Some small peaks and dips in the energy spectra, as well as the sudden drop around 140 MIP, are attributed to ADC saturation in various types of channels.

\begin{figure}[h!]
    \centering
    \includegraphics[width=0.49\linewidth]{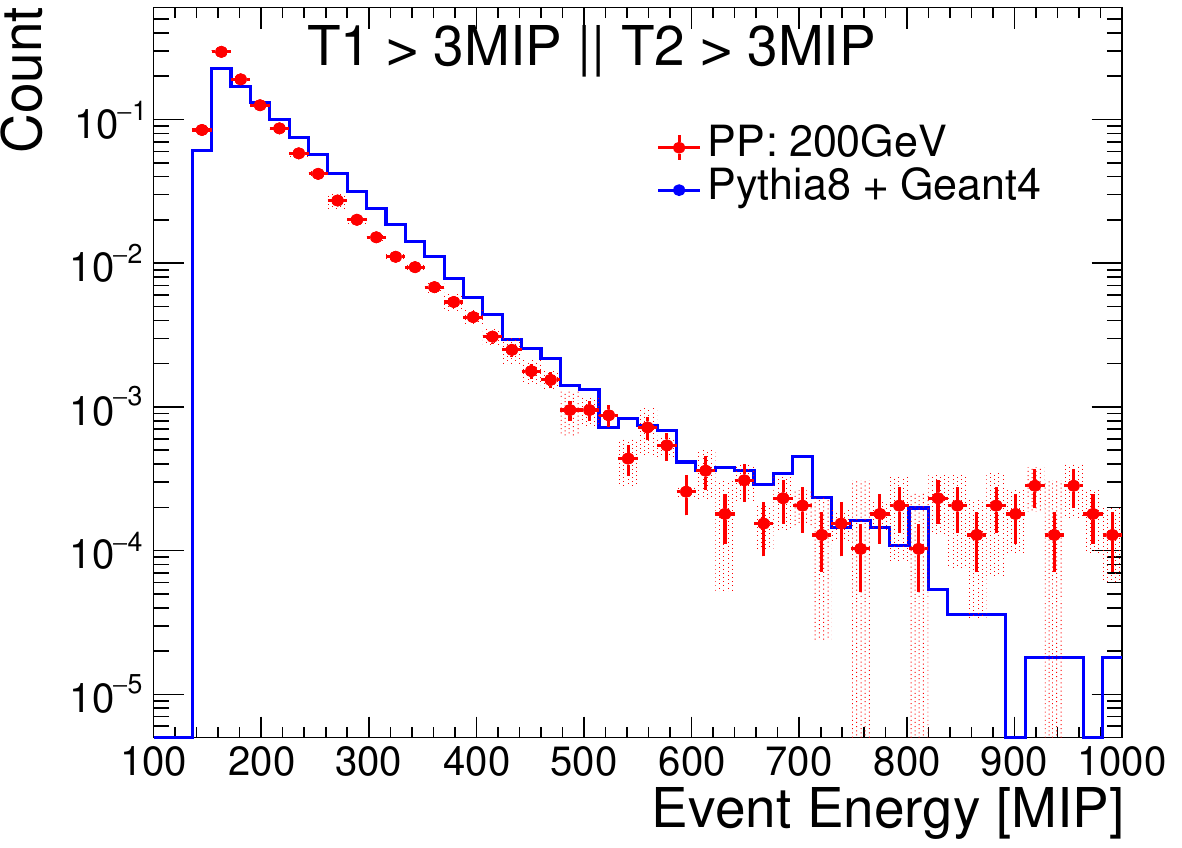}
    \includegraphics[width=0.49\linewidth]{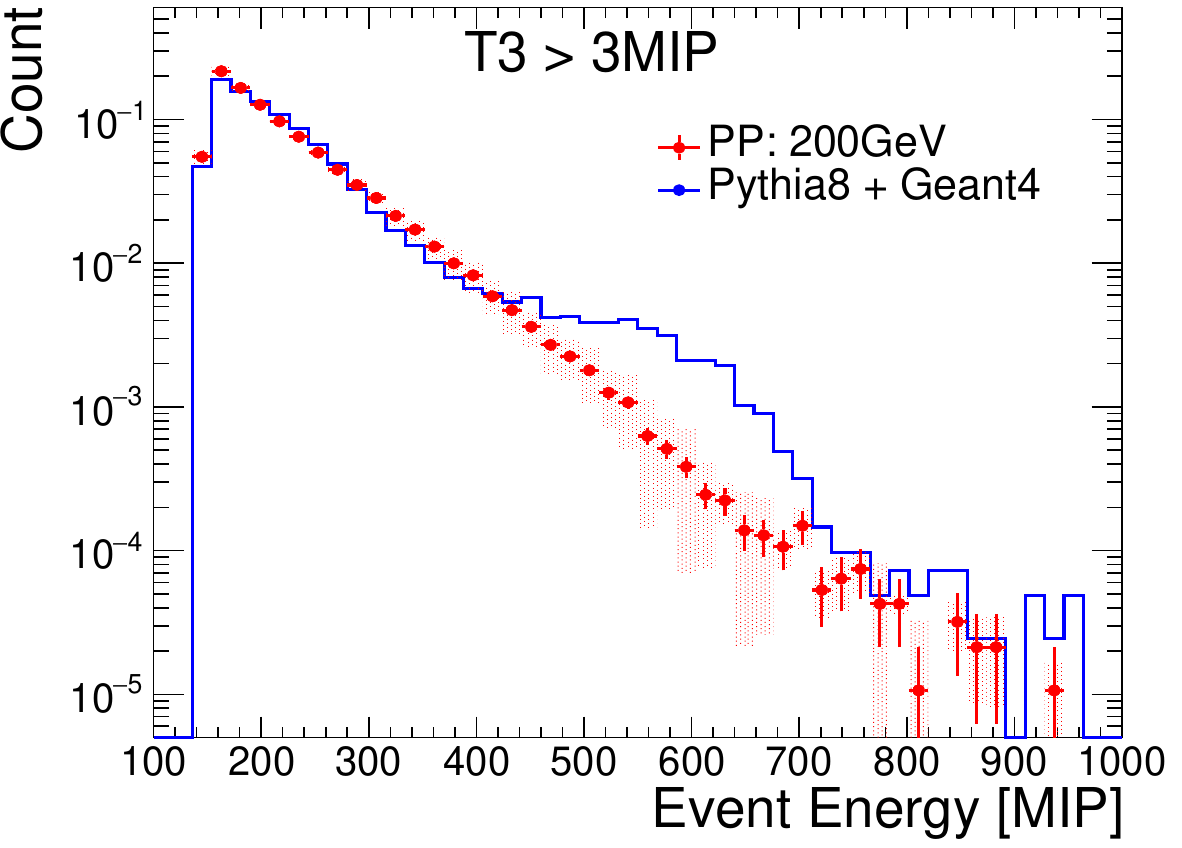}
    \caption{Comparison of data and simulation for event energy spectra for the T1$\vee$T2 trigger (left) and the T3 trigger (right). All distributions are normalized to unity. Error bars and bands represent statistical and systematic uncertainties, respectively.}
    \label{fig:eventenergy-comparison}
\end{figure}

Figure~\ref{fig:eventenergy-comparison} shows the event energy spectra for T1$\vee$T2 and T3 events. The data spectra for the T1$\vee$T2 events exhibit a steeper distribution. The simulation describes the spectra for T1$\vee$T2 reasonably well, although it predicts a feature in the 400–700 MIP range that is not observed in the data. At high energies, the T1$\vee$T2 data spectra show a flattening beyond 700 MIP, possibly indicating a background source, such as beam-gas interactions or another origin. These potential backgrounds are not accounted for in the simulation. In the T3 data, statistical uncertainties in this region prevent us from determining whether the same pattern is present.

\begin{figure}[h!]
    \centering
    \includegraphics[width=0.49\linewidth]{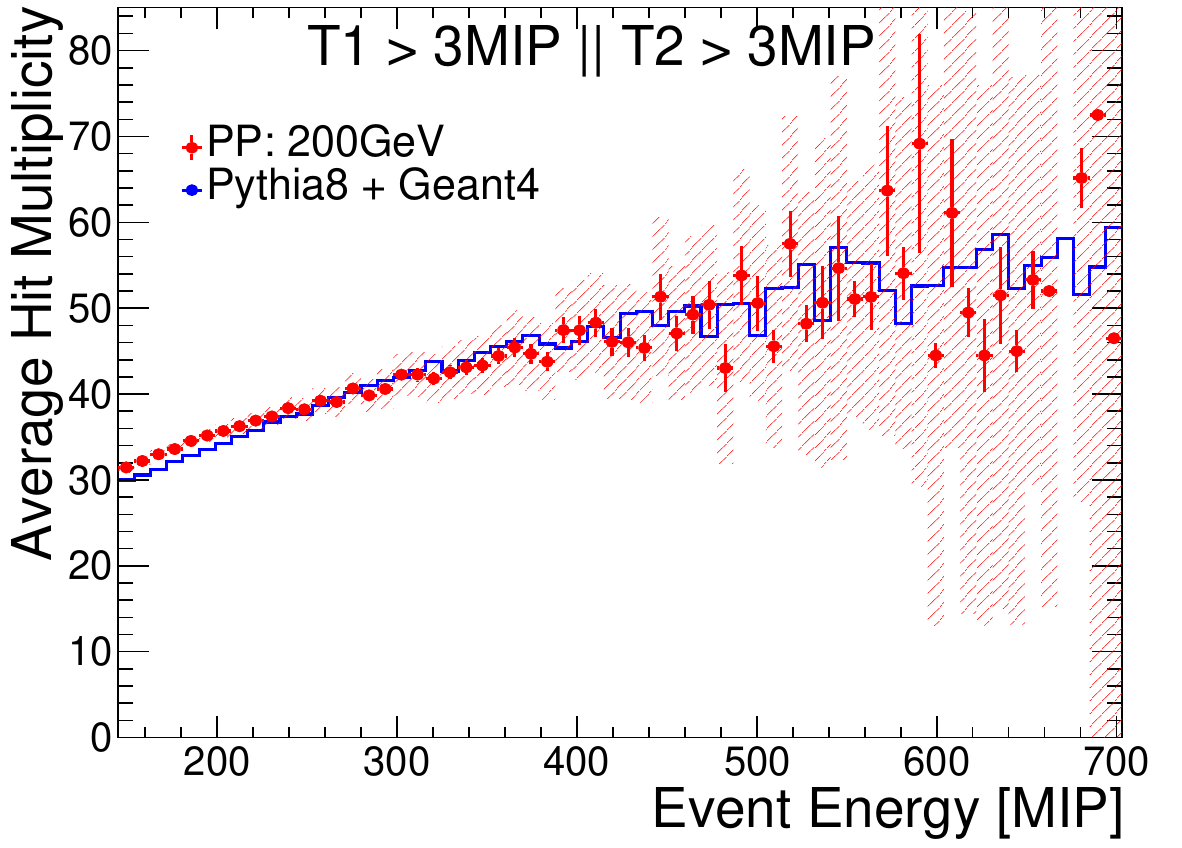}
    \includegraphics[width=0.49\linewidth]{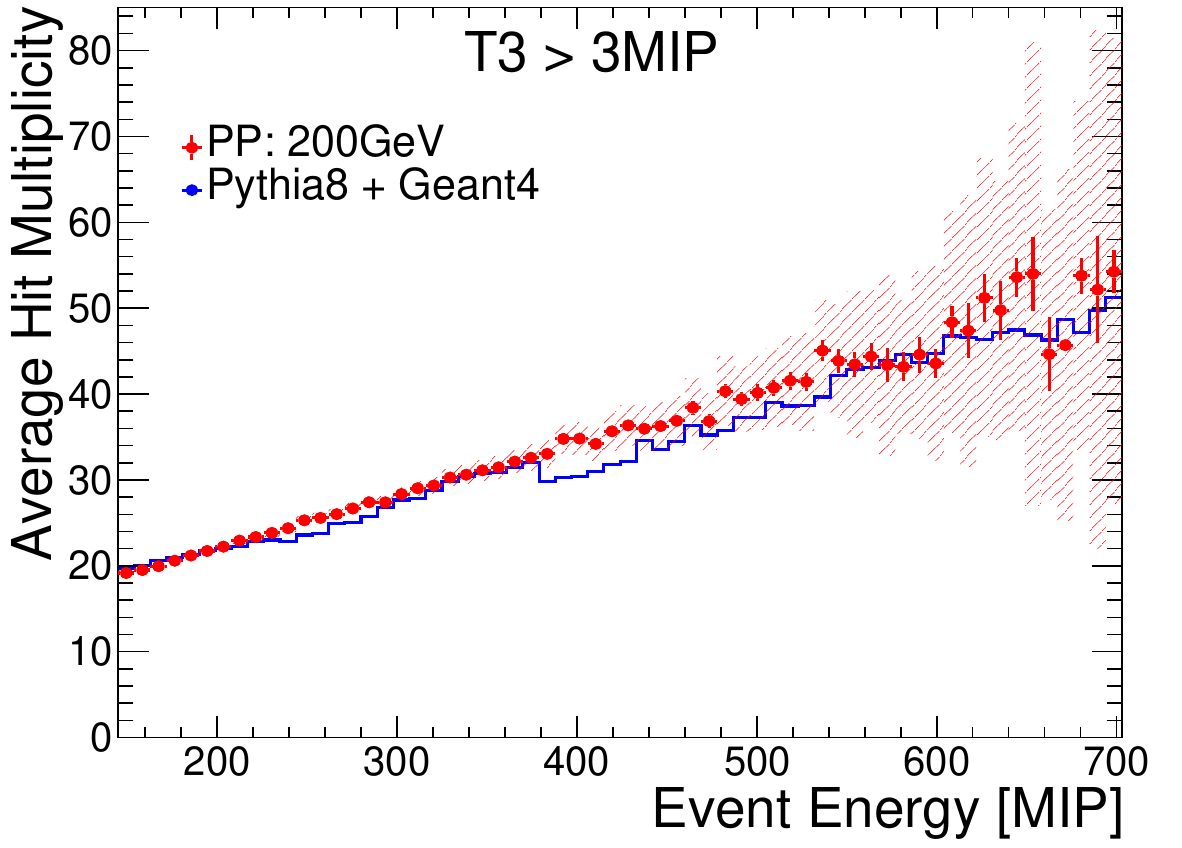}
    \caption{Comparison of hit multiplicity vs.~event energy between data and simulation. Error bars and bands represent statistical and systematic uncertainties, respectively.}
    \label{fig:event-energy-vs-hit-multiplicity}
\end{figure}

Figure~\ref{fig:event-energy-vs-hit-multiplicity} shows the relationship between hit multiplicity and event energy. A linear dependence is observed in both cases. The T1$\vee$T2 events exhibit higher overall multiplicity compared to T3 events. The simulation describes the trends for both triggers.

%% file: conclusions.tex
\section{Summary and Outlook}
\label{sec:conclusions}
We reported the first installation, commissioning, calibration, and long-term operation of a SiPM-on-tile calorimeter in a collider. This milestone was achieved during the 2024 RHIC run with 200 GeV proton-proton collisions. Using the dark current of a reference SiPM, the fluence to which this prototype was exposed is about $10^{10}$ 1 MeV $n_{\text{eq}}/\text{cm}^2$. This fluence exceeds the annual exposure expected for most of the forward hadronic calorimeter at the EIC at maximum design luminosity, thus providing a critical proof of concept for its future operation.

A data-driven calibration strategy was used, exploiting signals from minimum-ionizing particles during proton-proton collisions and an intercalibration of a dual-range ADC readout. The calibration was performed channel-by-channel and at various times throughout the run. While the increased dark current led to an increase in noise, reflected in pedestal width increases, the MIP calibration remained feasible. The MIP peak was always at least three times the pedestal width and remained stable over time within 10$\%$.

These results indicate that the system performance of the SiPM-on-tile calorimeter, with the current generation of SiPMs, can easily sustain $10^{10}$ 1 MeV $n_{\text{eq}}/\text{cm}^2$ without cooling and while operating at room temperature. Since the pedestal width scales with the square root of the current, which is proportional to fluence, extrapolating these results suggests that the same procedure could be performed at $10^{11}$ 1 MeV $n_{\text{eq}}/\text{cm}^2$ without significant difficulty. However, at $10^{12}$ 1 MeV $n_{\text{eq}}/\text{cm}^2$, calibrating the prototype becomes challenging, except for the 3 mm SiPM hexagonal-cell channels, which represent the configuration that maximizes the signal-to-noise ratio.

A \textsc{Geant4} simulation, based on the \textsc{Pythia8} Monte Carlo generator, describes the data reasonably well for the hit spectra, hit multiplicity, and event energy spectra for two trigger topologies. This further demonstrates that detector simulations are well understood and complement single-particle beam tests.

This successful proton-proton run will continue during the 2025 RHIC gold-gold run, extending until the final day of RHIC operation. Building on these promising results, we plan to upgrade the prototype by increasing the number of readout channels and integrating the CALOROC ASIC chip, specifically designed for ePIC calorimeters. 

The lessons learned from this experimental campaign, including calibration and monitoring, will guide the operational strategies and design of subdetectors at the EIC, particularly those expected to operate under the highest radiation levels, such as the CALI (insert of the forward hadronic calorimeter) and the Zero-Degree Calorimeter.